\documentclass{article}

\usepackage{arxiv}

\usepackage[utf8]{inputenc} 
\usepackage[T1]{fontenc}    
\usepackage{hyperref}       
\usepackage{url}            
\usepackage{booktabs}       
\usepackage{amsfonts}       
\usepackage{nicefrac}       
\usepackage{microtype}      
\usepackage{lipsum}		
\usepackage{graphicx}
\usepackage{natbib}
\usepackage{doi}

\usepackage{booktabs}   
\usepackage{multirow}   

\usepackage[toc,page]{appendix}

\title{Sea ice floe segmentation in close-range optical imagery using active contour and foundation models}

\date{December 8, 2025}	

\author{Giulio Passerotti \\
	    The University of Melbourne, Australia\\
	    \texttt{giuliopasserotti@gmail.com} \\
	    \And
    	Alberto Alberello \\
    	University of East Anglia, Norwich, UK
    	\And
    	Marcello Vichi \\
    	University of Cape Town, South Africa
    	\And
    	Luke G.~Bennetts \\
    	The University of Melbourne, Australia
        \And
        James Bailey \\
        The University of Melbourne, Australia
        \And
        Alessandro Toffoli \\
        The University of Melbourne, Australia \\
        \texttt{toffoli.alessandro@gmail.com}}

\renewcommand{\shorttitle}{}

\hypersetup{
pdftitle={Sea ice floe segmentation in close-range optical imagery using active contour and foundation models},
pdfsubject={physics.ao-ph},
pdfauthor={Giulio Passerotti},
pdfkeywords={sea ice, ice floe identification, computer vision, image segmentation, active contour model, foundation model, segment anything model},
}

\begin{document}
\maketitle

\begin{abstract}
The size of sea ice floes in the marginal ice zone (MIZ) is a key factor influencing ice coverage, albedo, wave propagation, and ocean--atmosphere energy exchanges. Floe size can be observed by processing visual-range imagery from ships, aircraft, or satellites. However, autonomously capturing floe boundaries remains challenging, particularly due to sea ice heterogeneity, which impairs boundary definition and reduces image clarity. This study evaluates the accuracy of sea ice floe segmentation using the gradient vector flow (GVF) active contour method, the deep learning-based Segment Anything Model (SAM), and a hybrid approach combining GVF and SAM. Methods are evaluated on a representative subset of a large dataset of close-range, high-resolution imagery collected from cameras aboard an icebreaker during an Antarctic winter expedition. Spanning a wide range of ice conditions and image clarity in the MIZ, the subset provides a rigorous segmentation test bed. Performance is assessed in terms of floe detection accuracy, size distribution, and ice concentration, with results compared against a manually segmented benchmark. Results indicate SAM, in prompt-driven mode, offers the best balance between accuracy and computational efficiency. Its strong performance in estimating sea ice concentration and detecting floes, while maintaining close agreement with benchmark floe size distributions, makes it suitable for real-time applications and scalable analyses of large imagery datasets. Compared with SAM, the combined SAM-GVF method provides more accurate floe boundary delineation, although at much higher computational cost, and is therefore better suited for analyses requiring precise floe shapes.
\end{abstract}


\section{Introduction}
\label{sec:introduction}

Sea ice modulates the interaction between the atmosphere and ocean at high latitudes \citep[e.g.,][]{landwehr2021esd,Bennetts_SORev_2024}. Its role in the marginal ice zone (MIZ), a dynamic transition region between open ocean and pack ice, is particularly complex and not yet fully understood due to sea ice heterogeneity \citep[e.g.,][]{alberello2022three,tersigni2023high}. The MIZ is characterized by a mix of fragmented, highly mobile sea ice pieces called floes \citep{alberello2022three,vichi2022indicator,tersigni2023high}. These floes vary widely in type, concentration, and size and are continuously reshaped by ocean currents, waves, and atmospheric and thermodynamic processes \citep[e.g.,][]{herman_sizes_2021,tersigni2023high}. The fragmentation of sea ice governs how waves transmit energy through it and, concurrently, how ice further breaks up under their forcing \citep{dolatshah2018hydroelastic,montiel2016attenuation,meylan_floe_2021,passerotti2022interactions,day2024}. It also influences the ocean surface energy budget by modulating heat fluxes \citep{andreas2010parametrizing,bourassa2013high,tersigni2023high}, thereby affecting melting processes \citep[e.g.,][]{horvat2016interaction}. Moreover, the floe size distribution influences sea ice rheology, controlling ice deformation and movement under external forces. This, in turn, affects overall ice dynamics, including the formation of ridges, leads (narrow openings in sea ice), and regions of open water \citep{feltham2005granular,alberello2020drift,herman_sizes_2021,womack2022atmospheric,womack2024contrast}, adding further complexity to the MIZ \citep{tersigni2023high}.

Accurate observations of floe geometry, coverage, size, and dynamics are key to advancing and validating modern sea ice models \citep[e.g.,][]{bennetts2017impacts,roach2018emergent,wang2023tc,buckley2024seasonal,day2024}, as these metrics strongly influence sea ice evolution and related climate feedbacks. Large scale observational datasets capturing spatial and temporal variability across the MIZ are necessary to constrain floe size distribution models, which are sensitive to assumptions about floe geometry and evolution \citep{Bateson2022,Horvat2022}. Floe characterization has long relied on high-resolution imagery combined with image processing techniques, and is now routinely extended using satellite, airborne, and shipborne sensors \citep[e.g.,][]{alberello2022three,alsharay2022improved,herman_sizes_2021,lu2016shipborne,paget2001determining,parmiggiani2019image,sandru2020complete,toyota2006characteristics,zhang2023ice,zhang2014image,weissling2009eiscam,Hwang2017}. Manual approaches, such as hand-drawn floe delineation, can be accurate, but they are painstaking and time‑intensive, and do not scale to the number of images required for robust statistical analyses \citep{Horvat2022}. Classical automated methods, such as threshold segmentation, rely on user-defined values, which can introduce inconsistencies due to variations in sea ice conditions across the image \citep{heyn2017system, lu2008aerial, alberello2019brief}. To address this limitation, more advanced techniques, including k-means clustering \citep{weissling2009eiscam, lu2016shipborne}, the gradient vector flow (GVF) snake algorithm \citep{zhang2014image}, and watershed segmentation \citep{dumas2021aerial, parmiggiani2019image}, have been applied to improve automation and robustness in sea ice analysis. Despite their advantages, these methods face challenges when applied to complex atmospheric and ice conditions that reduce the clarity of sea ice imagery. Although they perform well under optimal conditions, accuracy and consistency decline markedly in images with reduced clarity. Factors such as coarse resolution, cloud cover, fog, variable lighting, and low contrast between ice and water reduce image clarity. Additionally, snow cover and the presence of interstitial ice, thin ice forming in the spaces between floes, can obscure floe boundaries, creating the illusion of a more uniform surface or larger floes. These challenges complicate the identification of individual floes, particularly when they are densely packed, and make automated analysis of large image datasets problematic.

In recent years, deep learning has significantly advanced image processing, demonstrating high accuracy in object detection, classification, and segmentation across domains such as medical imaging, autonomous vehicles, and environmental monitoring. Convolutional neural networks (CNNs) are commonly used to classify pixels based on patterns learned from training data. In the context of sea ice, CNNs have been applied to differentiate elements such as ice, water, and sky \citep{zhang2022semantic, panchi2021supplementing, dowden2020sea}, prioritizing semantic segmentation rather than distinguishing individual ice floe properties. Applications specifically aimed at floe characterization have shown promising results \citep[e.g.,][]{zhang2023ice}, though over-segmentation remains a challenge, particularly in more heterogeneous sea ice conditions. This is partly due to the reliance of deep learning models on large labeled datasets \citep{maska_2023_wve5d-2gn32}. In this respect, the lack of sufficiently diverse training data that accurately captures the complexity and variability of sea ice remains a significant hurdle.

Vision foundation models have emerged as a promising alternative for image segmentation tasks \citep{Dosovitskiy2020anImage}. These large-scale deep learning models are pre-trained on massive datasets and can be adapted to new tasks with minimal additional training, making them particularly suitable for applications where labeled data is scarce \citep{raghu2021do}. One notable example is the Segment Anything Model (SAM), developed by Meta AI Research \citep{Kirillov2023segment}. Trained on an extensive dataset of one billion masks across 11 million images, SAM can generate segmentation masks without requiring additional labeled data. Its potential for ice observations was demonstrated by \citet{shankar2023semantic}, who applied SAM to segment large icebergs in satellite images, achieving accuracy comparable to specialized CNN models. Nevertheless, SAM was tested on a limited imagery dataset, and its performance in more challenging conditions remains unclear.

Here, we use an image acquisition protocol deployed on the icebreaker S.A.\ Agulhas II during expeditions to the Antarctic MIZ in the austral winter \citep{alberello2019brief, alberello2022three, toffolietal_prlrogue_2024}. The resulting dataset, the most extensive currently available, comprises high-resolution, close-range images of sea ice that offer detailed spatial and temporal coverage, capturing a broad range of ice conditions under varying levels of clarity. Taking advantage of this diversity, we assess the robustness of different segmentation approaches, namely the traditional active contours model (GVF Snake algorithm), the advanced SAM deep learning model, and a novel integration of the two. To support this evaluation, we develop a fully automated processing pipeline that enables their systematic and autonomous application across the varied ice conditions represented in the dataset. We conduct detailed, floe-by-floe visual comparisons against manually segmented benchmarks for two representative scenes with contrasting ice conditions, revealing the segmentation discrepancies, strengths, and limitations of each method. Beyond qualitative assessment, we perform a statistical evaluation using a benchmark of over 2,800 manually segmented floes, the most extensive dataset of its kind for sea ice segmentation analysis, to identify the most effective method for detecting ice floes and estimating floe size and ice concentration under diverse conditions.

\section{Image acquisition}
\label{sec:images_acquisition}

\begin{figure}
\centering\includegraphics[width=0.9\textwidth]{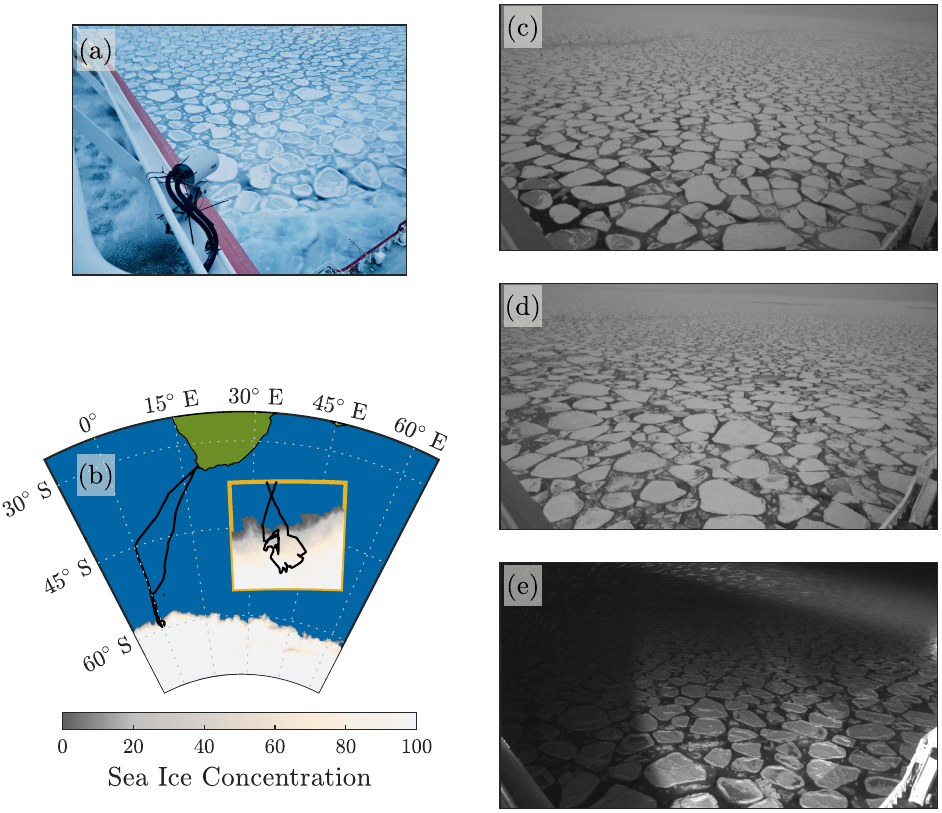}
\caption{Antarctic SCALE expedition, winter 2022: (a)~camera mounted on S.A.~Agulhas II; (b)~ship track from departure on 11~July 2022 to return on 31~July 2022, entering the sea ice region on 19~July and exiting on 24~July, overlaid on sea ice concentration from the University of Bremen AMSR2 ASI product \citep[6.25\,km grid;][]{spreen2008sea}; (c--e)~high-resolution images of the sea ice surface captured under (c)~conditions featuring distinct ice floes with defined edges, (d)~diffuse sea ice boundaries characterized by densely packed, irregularly shaped floes with indistinct or visually merged edges, and (e)~at nighttime.}
\label{fig:cam_set}
\end{figure}

Close-range optical sea ice imagery was acquired underway from a sensor installed on the icebreaker S.A.~Agulhas~II during austral winter expeditions to the Antarctic MIZ in the Eastern Weddell Sea (Fig.~\ref{fig:cam_set}). These expeditions were part of the Southern oCean seAsonaL Experiment (SCALE). For this study, we focus on data collected during the July--August 2022 expedition \citep{vichi_2023_7902557}. The MIZ was reached at approximately 58$^{\circ}$S and 1$^{\circ}$W on 19~July 2022 (Fig.~\ref{fig:cam_set}b). The vessel then continued southward until it reached consolidated pack ice at latitude 58.8$^{\circ}$S ($\approx$85\,km from the ice edge), marking the inner boundary of the MIZ. Overall, the expedition spent six days in the MIZ before heading back North.

The sensor consisted of a GigE monochrome industrial camera (Fig.~\ref{fig:cam_set}a) equipped with 2/3'' Sony CMOS Pregius sensor and 5\,mm F1.8 C-mount lenses (angle of view $\approx$120$^{\circ}$). It was mounted on the port side of the monkey bridge at 25\,m above the water line and tilted by $\approx$70$^{\circ}$ relative to the ground. The field of view covered a portion of the ocean surface of $\approx$ 200$\times$200\,m$^2$. Images were recorded continuously with a resolution of 1920$\times$1080 pixels and at a sampling rate of 1\,Hz. The sensor was paired with a synchronized inertial measurement unit (IMU) to record variations in the camera’s position relative to the horizontal plane during navigation.

Around 90,000 images were collected during the expedition, covering both day and night, with nighttime imaging supported by flashlights illuminating the sea ice (see sample images in Fig.~\ref{fig:cam_set}c--e). Consequently, the database spans a wide range of ice and atmospheric conditions, with approximately 70\% capturing complex sea ice scenarios. These situations are characterized by low light (Fig.~\ref{fig:cam_set}e), densely packed and irregularly shaped floes, and diffuse boundaries, where floe edges are indistinct or visually merge due to snow cover, interstitial ice, or low contrast (Fig.~\ref{fig:cam_set}d).

\section{Pre-processing}

\subsection{Orthorectification}
\label{sec:persp_correction}

The tilt of the sensor causes perspective distortion, making ice floes closer to the camera appear larger than those farther away. This distortion creates inconsistencies in scaling, which vary across each individual image and compromise the accuracy of the conversion between pixel measurements and real-world distances. To address this issue, a correction process known as orthorectification was applied \citep{luhmann2023close}. This process rectifies the sensor orientation by adjusting the distorted pixels as if the image were projected onto a plane perpendicular to the optical axis.

Orthorectification interpolates over the pixels in the original image based on the camera projection matrix, which maps the conversion of a three-dimensional point cloud in the real world to a two-dimensional plane (the image) through intrinsic and extrinsic parameters \citep{hartley2003multiple}.  The former depends on how the sensor captures images and include, but are not limited to, the focal length, aperture, resolution, and the optical center of the camera’s sensor. These parameters are derived from a calibration process \citep[e.g.,][]{zhang2000flexible}, which evaluates image distortion by analyzing multiple images of a planar pattern with known geometry \citep[e.g., a chessboard; see][]{sturm1999plane,bouguet2004camera}. Using the method in \citet{zhang2000flexible}, multiple images of a chessboard are captured from various orientations and positions across the field of view. The algorithm automatically detects pattern corners and estimated the intrinsic parameters by fitting the observed geometry to its known dimensions, while correcting for radial and tangential lens distortions. Potential errors, such as imperfect corner localization or slight non-planarity, are minimized by employing a rigid board, good lighting, and a wide range of viewing angles. Parameter drift due to changes in focus or zoom is avoided by keeping lens settings fixed between calibration and deployment. Calibration was performed before and after the expedition to verify that intrinsic features remained stable throughout the campaign.

The extrinsic parameters, on the other hand, describe the translations and rotations of the sensor relative to a reference Cartesian coordinate system. For simplicity, the camera was positioned at the origin of the horizontal axis, so translations in the $x$ and $y$ directions were zero, and at a height of 25\,m above the water line. The rotations involve three components: tilt around the transverse axis, which was imposed during deployment; roll around the optical axis, which was minimized during installation to ensure alignment with the horizon; and heading around the vertical axis, which aligned with the sensor’s direction. The latter is relevant only when targeting a specific object within the field of view, but becomes negligible when the target is an extended area (such as the ocean surface in this case). In this analysis, translations and rotations are assumed to remain constant over time, effectively neglecting the influence of wave-induced ship motion. This approach aligns with similar applications on mobile platforms, such as aircraft \citep{zhang2014image,parmiggiani2019image} and ships \citep{heyn2017system,alberello2019brief,alberello2022three,tersigni2023high}. A sensitivity analysis addressing ship motion and validating this assumption is discussed in \S\ref{shipmotion}.

\begin{figure}
\centering\includegraphics[width=0.85\textwidth]{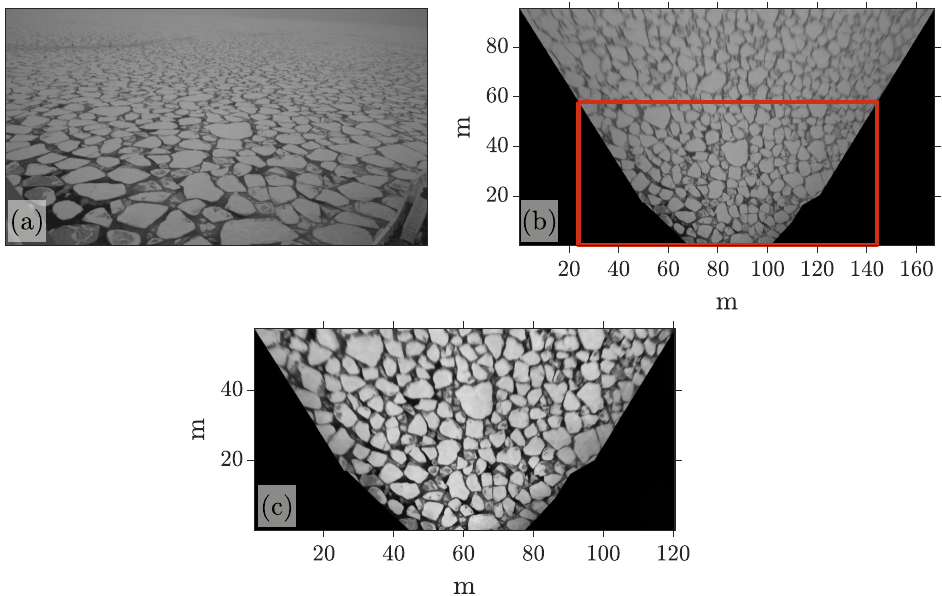}
\caption{Sample sea ice image (a) from the Antarctic SCALE expedition, winter 2022, after perspective correction, highlighting the region of interest for analysis in red (b). The selected region after image enhancement, including gray-level homogenization through Gaussian bilateral filtering, anisotropic diffusion, and contrast enhancement using CLAHE (c).}
\label{fig:persp_image}
\end{figure}

Orthorectification is performed using the CameraTransform Python package \citep{gerum2019cameratransform}, while calibration is carried out with MATLAB’s Computer Vision Toolbox \citep{bouguet2004camera}. The process focused on a rectangular portion of the ocean surface, approximately 95\,m $\times$ 165\,m, selected based on a visual assessment to include clearly visible floes. The resolution (i.e., pixel-to-metre conversion) was set to 0.05\,m to ensure both high object accuracy and the ability to capture small-scale sea ice features \citep[in the order of tens of centimetres;][]{alberello2019brief}. A sample image and its orthorectified counterpart are shown in Fig.~\ref{fig:persp_image}a,b.

Floe edges lose sharpness with distance from the sensor (see upper band of Fig.~\ref{fig:persp_image}b). To avoid ambiguity in the far field and, hence, reduce the risk of detecting unrealistic floes, the workable portion of the orthorectified images is determined by qualitatively assessing multiple images to identify where floe edges noticeably degrade. This effectively restricts the analysis to an area of about 60\,m $\times$ 120\,m close to the sensor (see area within the red rectangle in Fig.~\ref{fig:persp_image}b).

\subsection{Effect of ship motion}
\label{shipmotion}

\begin{figure}
\centering\includegraphics[width=0.85\textwidth]{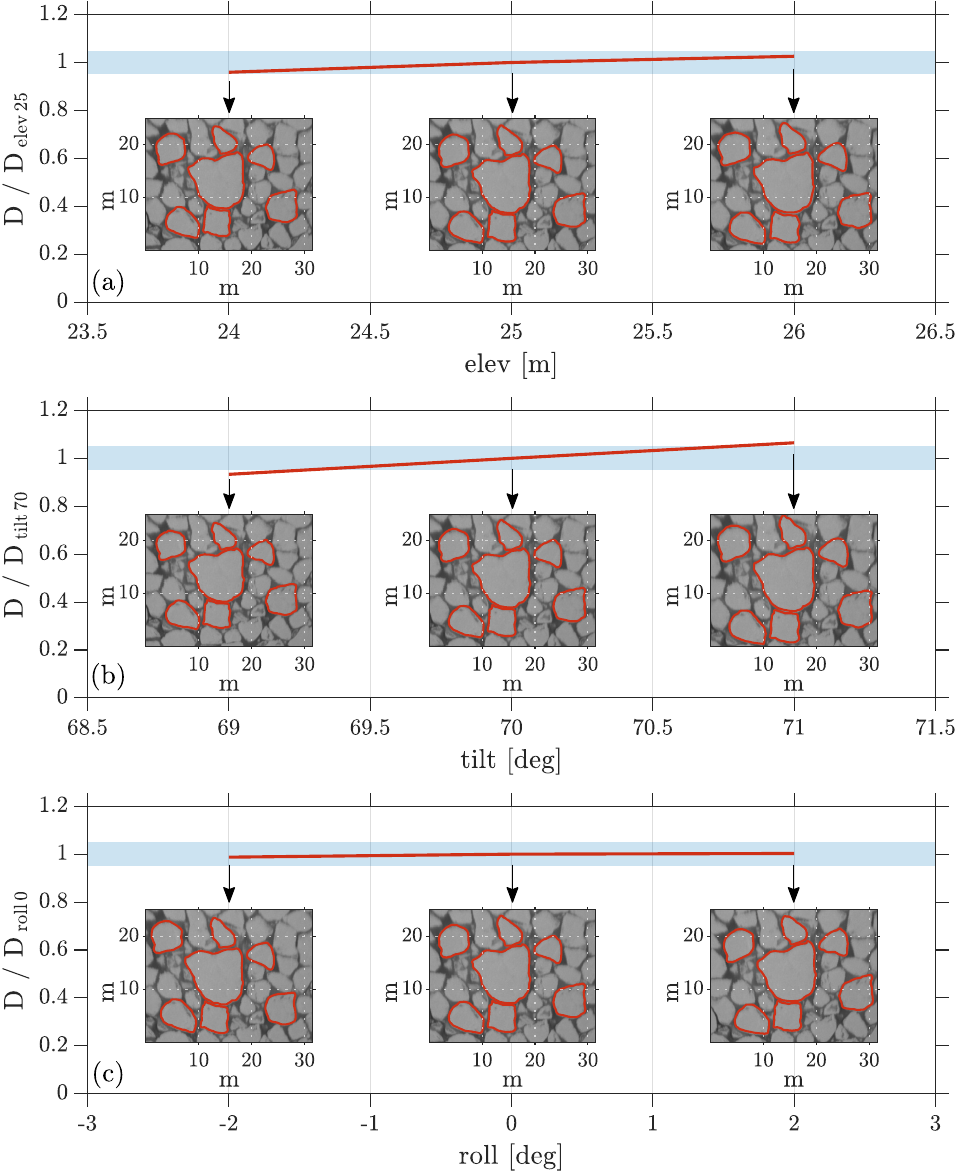}
\caption{Uncertainty analysis of extrinsic camera parameters on average floe diameter, with the floes used in the analysis highlighted in red. Variations include (a)~$\pm$ 1\,m elevation, (b)~$\pm$ 1$^{\circ}$ tilt, and (c)~$\pm$ 2$^{\circ}$ roll. Floe diameters are normalized against values derived from orthorectification using ship-measured extrinsics. The shaded blue area indicates a 5\% uncertainty range.}
\label{fig:uncert}
\end{figure}

The assumption that the sensor remains fixed relative to the ocean surface holds true in calm seas but does not apply in open ocean conditions \citep[cf.][]{alberello2022three}. Movement of the supporting platform can alter the camera’s orientation affecting orthorectification and floe size dimensions. To assess sensitivity to ship motion, an image captured in clear sky conditions was used as a proxy. The orthorectification was evaluated by artificially introducing misalignments of $\pm$1\,m in elevation, $\pm$1$^{\circ}$ in tilt, and $\pm$2$^{\circ}$ in roll relative to the reference configuration established during calm seas. These extreme values of ship motion are determined based on direct measurements obtained from the IMU positioned in close proximity to the camera during the expedition. A few clearly distinguishable floes were manually segmented after orthorectification by outlining their contours using photo editing software for analysis (see Fig.~\ref{fig:uncert}).

Figure~\ref{fig:uncert} shows the average variation in floe size, quantified by the diameter relative to the dimension extrapolated from the image with the sensor in its original position, as a function of misalignment. Uncertainties in floe size due to elevation variations are confined to within 5\% (Fig.~\ref{fig:uncert}a). Tilt variations have a slightly greater impact (Fig.~\ref{fig:uncert}b), but the effect remains just over 5\%. Roll variations have the least impact, with floe size changes under 2\% (Fig.~\ref{fig:uncert}c). It is important to note that the impact of ship motion is assessed by considering elevation, tilt, and roll independently. However, the ship's design and control systems are engineered to minimize the chance that these motions peak simultaneously \citep{fossen2011handbook}. Consequently, their combined effect is typically lower than the sum of their individual extremes, supporting the assumption that the camera's motion can be neglected.

\subsection{Image enhancement}
\label{sec:enhancement}

Sea ice imagery captured in natural settings is influenced by varying light, contrast, shadows, and impurities (see Fig.~\ref{fig:cam_set}c--e). Consequently, ice floe boundaries in the raw images are often unclear, making automated segmentation challenging \citep[cf.][]{maska_2023_wve5d-2gn32}. To address this, image enhancement techniques are applied to sharpen the edges by homogenizing gray-scale levels \citep[e.g.,][]{parmiggiani2019image,alberello2019brief}. It is accomplished by combining an edge-preserving Gaussian bilateral filter \citep{tomasi1998bilateral} and an anisotropic diffusion filter \citep{perona1990scale}, which smooths regions with low gradients while preserving ice floe boundaries. Further enhancement is achieved by improving contrast with contrast-limited adaptive histogram equalization (CLAHE) \citep{zuiderveld1994contrast}, which accentuates boundaries and reveals fine details. Fig.~\ref{fig:persp_image}b,c illustrates the transition from the original to the enhanced image. Notably, the adaptive nature of this method allows it to tailor adjustments to each image’s specific characteristics, eliminating the need for predefined thresholds.

For night photographs, enhancement was limited to areas illuminated by the vessel’s spotlight. These regions are identified using haze removal techniques \citep{he2010single}, which estimate the illumination map by inverting the low-light image, applying dehazing, and reinverting the result. This highlights the spotlight-illuminated areas for further analysis. As with daytime images, segmentation performance in these illuminated areas is influenced by sea ice complexity.

\section{Image segmentation}

\subsection{Active contours model (GVF)}
\label{s_gvf}

Active contour models (otherwise known as snakes) adjust an evolving curve to align with floe boundaries \citep{kass1988snakes}. Poor initialization of the contour can lead to uncertain boundaries, while contours set close to the true edges enable faster and more precise convergence \citep{mcinerney1996deformable}. To eliminate the need for manual initialization, an automated contour generator is introduced. This tool adapts to various floe shapes, making the segmentation process fully autonomous.

\begin{figure}
\centering\includegraphics[width=0.9\textwidth]{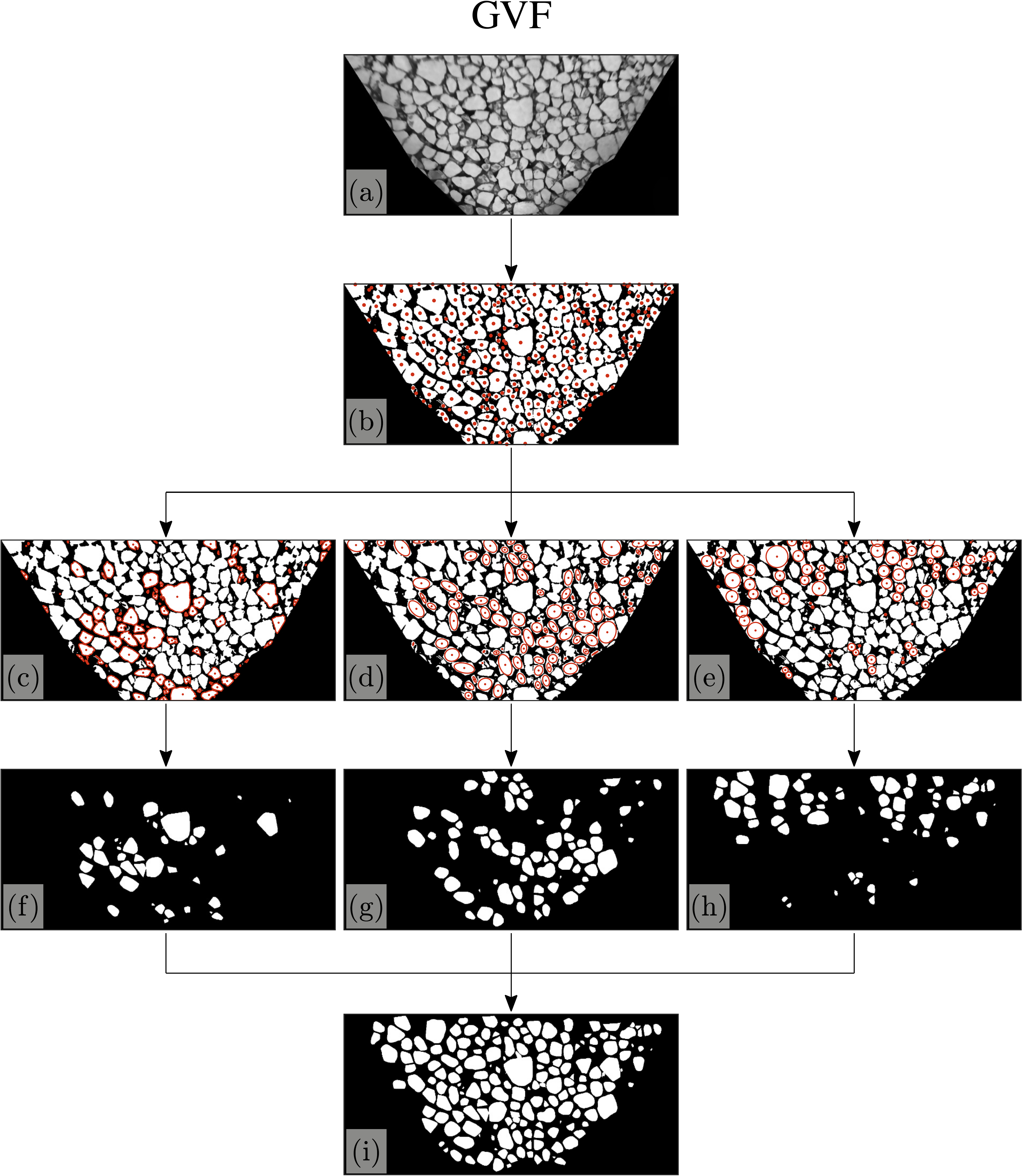}
\caption{Illustration of the GVF Snake algorithm workflow: (a)~enhanced sea ice image; (b)~seeds (red dots) representing floe centers on the binarized image; (c)~floe perimeters (red lines) identified from isolated, single-seed objects; (d)~elliptical initial contours (red lines) of floes containing a single seed after morphological erosion; (e)~circular initial contours (red lines) for remaining unused seed; (f)~floes identified directly from the binarization process (as shown in Fig.~\ref{fig:gvf_outputs}c); (g)~floes detected by the GVF snake algorithm initialized from elliptical contours; (h)~floes detected by GVF initialized from circular contours; (i)~final binary segmentation with all detected floes.}
\label{fig:gvf_outputs}
\end{figure}

Fig.~\ref{fig:gvf_outputs} illustrates the phases in the image segmentation process. The process begins by separating the foreground (i.e., the objects) from the background, converting the enhanced ice image into a binary format (Fig.~\ref{fig:gvf_outputs}b). This is done by selecting an appropriate threshold from the gray-scale histogram of the entire image. Pixels below the threshold are labeled as non-feature pixels (background), which includes both the open water fraction and interstitial ice that lacks a specific gradient, and assigned a value of zero, while those above the threshold are labeled as feature pixels (foreground), representing floes, and assigned a value of unity (represented by the white islands in Fig.~\ref{fig:gvf_outputs}b). In cases with many objects and low gray-scale contrast, the histogram lacks clear modes, leading to ambiguities in selecting the threshold. To resolve this, the Otsu method \citep{otsu1979threshold} is employed for automated thresholding. The Otsu method optimally determines the threshold by maximizing the variance between foreground and background pixel intensities. To enhance accuracy, this binarization is applied to small clusters of $12\times{}12$\,m$^{2}$, each containing about 20 floes. The smaller cluster size results in more distinct modes in the gray-scale histogram, improving the effectiveness of the automated thresholding \citep{byun2021estimation}.

A distance transformation \citep{paglieroni1992distance,maurer2003linear} is applied to the binary image to compute a distance map, where each feature pixel is assigned a value corresponding to its distance from the nearest non-feature pixel. Local maxima in this distance map are pixels with values greater than or equal to those of all their immediate neighbors. These maxima serve as seeds, each marking the center of an individual object \citep{zhang2014image}. To improve accuracy, regional local maxima are considered by dividing the image into sub-regions, noting this approach was previously used for the binarization. If a sub-region is sharp, objects within it are well separated and typically have only one seed. Consequently, segmentation of these isolated objects follows directly from the binarization process. This initial phase of floe detection is illustrated in Fig.~\ref{fig:gvf_outputs}c,f, which shows the identified seeds and their corresponding segmented objects, respectively.

However, adjacent objects are merged through the process in most sub-regions, thus preventing segmentation through binarization alone. This results in large, irregular white regions containing multiple seeds (see Fig.~\ref{fig:gvf_outputs}b). In such cases, segmentation is performed using the gradient vector flow (GVF) Snake algorithm \citep{xu1998snakes}, which extends traditional active contour models by reducing dependence on the initial contour. This allows us to process objects with complex shapes efficiently, even in noisy or low-contrast images. Unlike traditional models, the initial contour shape does not constrain the final segmentation from adapting to irregular boundaries.

The adjacent objects are then separated by smoothing their edges through a morphological erosion process \citep{adams1993radial}. The newly formed objects are initialized using ellipses, which approximate the shape of ice floes \citep{alberello2019brief, passerotti2022interactions}. The initial contour is then chosen such that the second moment of the covariance matrix (a 2$\times$2 matrix that captures the object's spatial spread and orientation) aligns with that of the object \citep{mulchrone2004fitting} (see Fig.~\ref{fig:gvf_outputs}d).

In many cases, objects share significant portions of their edges with neighboring objects and cannot be separated successfully using morphological erosion. Instead, segmentation is performed by applying circles as initial contours to the seeds, as the circular shape is suitable for smaller floes \citep[e.g.,][]{alberello2019brief}. Using the distance transform, radii are chosen so that contours remain within the objects but are not too far from the presumed edges (Fig.~\ref{fig:gvf_outputs}e). Once initial contours were assigned to all seeds, the GVF snake algorithm was applied to expand the contours and identify the object’s boundaries (Fig.~\ref{fig:gvf_outputs}g,h).

The GVF snake algorithm can produce anomalies and inaccuracies, known as ``hallucinations'', where the segmentation detects features that do not correspond to actual floes. These occur when the algorithm misinterprets patterns or textures in the image---such as shadows, smooth gradients, or areas with low contrast---as floe boundaries. To identify and remove such spurious features, we evaluate the shape of each segmented object using coefficients of circularity and eccentricity, which quantify how closely contours resemble expected floe shapes. By applying deliberately high-valued thresholds, spurious features in the form of nearly perfect circles or line segments are removed (circularity \textgreater{}1 and eccentricity $\geq$0.9). These thresholds target segmentation noise across all images, such as isolated pixels, without affecting segmented floe masks. Additionally, objects with open boundaries at the image’s edges are excluded from the segmentation.

The final result is obtained by integrating outputs from each phase of the analysis (see Fig.~\ref{fig:gvf_outputs}i). On average, the computational time ranges from approximately 35\,s to 75\,s per floe, depending on floe morphology (i.e., shape, edge clarity, and surface texture), on a 2.3\,GHz Intel Core i9 8-core processor without GPU acceleration. Computational time is reported per floe, rather than per image, to avoid bias from variability in the number of floes per image. A summary of computational times is reported in Tables \ref{table1}, \ref{table2}, \ref{table3}, and \ref{table4}.

From the final binary mask, each ice floe is identified as a connected region of pixels using eight-neighborhood connectivity, meaning that two pixels are considered connected if they share an edge or a corner. Once the floes are identified, geometric properties such as area, perimeter, and diameter are derived. The area is determined by counting the total number of pixels in each region. The perimeter is obtained by tracing along the region boundary to measure its continuous outline \citep{biswas2018robust}. The diameter is calculated by equating the floe to a circle of the same area, thus representing the diameter of that hypothetical circle. The ice concentration is computed by summing the areas of all floes and dividing by the total image size excluding any masked-out, empty regions introduced by orthorectification. We note that ice concentration is used here only as a relative metric for comparison with the manual benchmark, which similarly excludes floes partially located at image edges. For absolute concentration estimates, partial floes at the image edges should be included.

\subsection{Segment Anything Model}

\begin{figure}
\centering\includegraphics[width=0.75\textwidth]{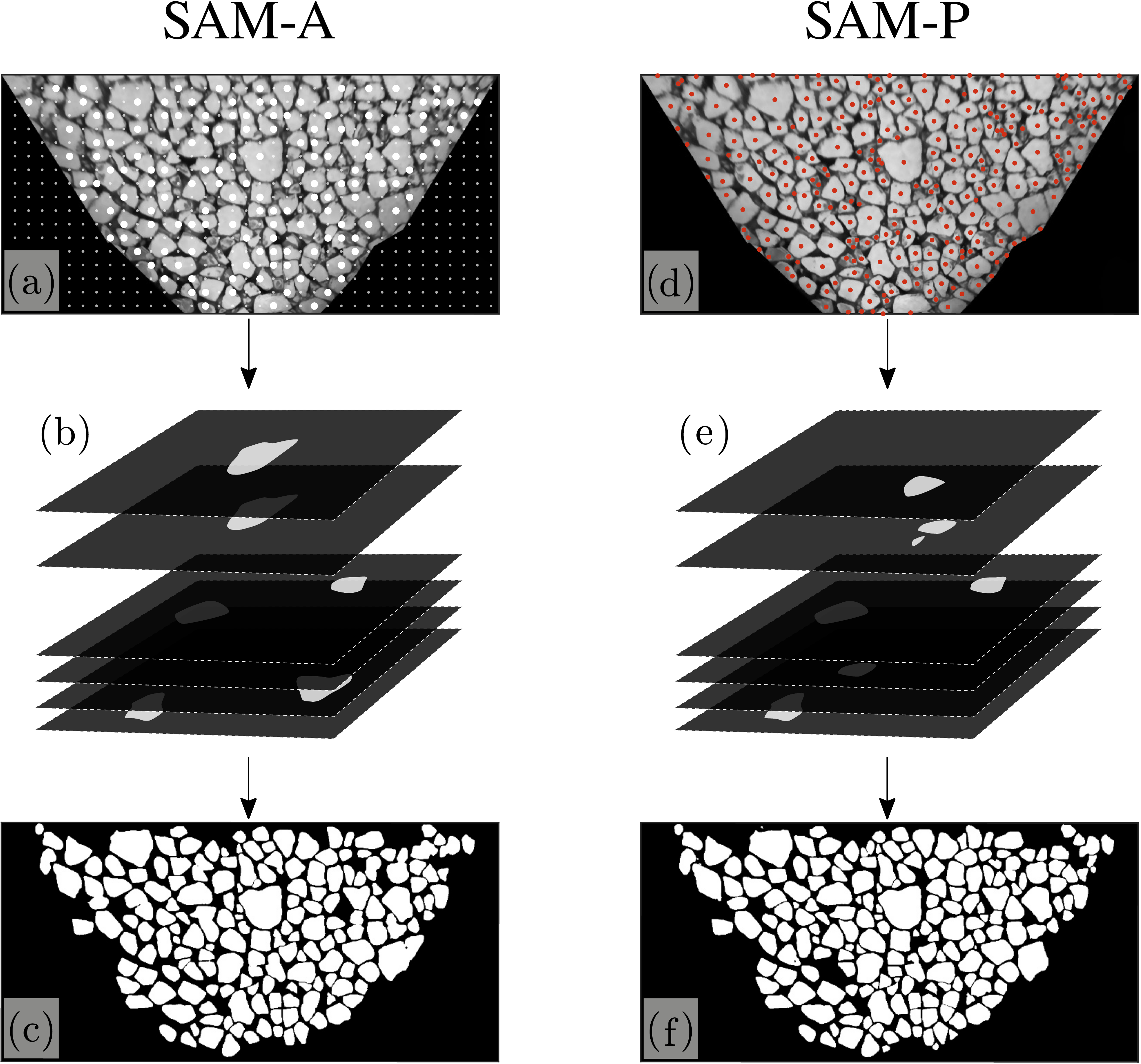}
\caption{SAM segmentation modes applied to enhanced sea ice imagery. (a--c)~Automatic segmentation mode (SAM-A): (a)~grid-based fixed points automatically selected by SAM; (b)~consolidation of masks returned by the model after filtering; (c) final unified binary mask of detected floes. (d--f)~Prompt-based segmentation mode (SAM-P): (d)~seed points (red dots) marking floe locations; (e)~consolidation of masks returned by the model after filtering; (f)~final unified binary mask of detected floes.}
\label{fig:samAuto_flow}
\end{figure}

Snake-based models rely on pixel-level features to segment images, where the features include smoothness, roughness, and contrast, which are specific to each image. In contrast, modern AI-driven approaches, exemplified by vision foundation models \citep{Dosovitskiy2020anImage}, leverage deep learning to identify general patterns from extensive training datasets. These models process entire images without prior knowledge of object shapes, thus eliminating the need for predefined contours or manual feature extraction.

The Segment Anything Model (SAM), developed by Meta AI Research \citep{Kirillov2023segment}, is a vision foundation model designed for general-purpose image segmentation. Built on the PyTorch deep learning framework \citep{ketkar2021introduction}, SAM operates in two distinct modes. The first (SAM-A) is an automatic mask generator that segments all identifiable elements in an image without user input, autonomously determining segmentation regions based on its internal parameters. The second mode (SAM-P) is prompt-based, allowing users to guide the segmentation process by specifying regions of interest using points or bounding boxes. This mode is particularly useful when precision and control over the segmented output are essential.

Tests on two representative images indicate that the newer version, SAM 2 \citep{ravi2024sam}, performs worse than the original SAM model for sea ice segmentation tasks, detecting on average 40\% fewer floes in automatic mode and 13\% fewer in prompt-based mode. A detailed comparison between SAM and SAM 2 is provided in \S\ref{sam2}. This finding contrasts with \citet{ravi2024sam}, who reported similar performance between the two versions, but aligns with \citet{sengupta2024sam}, who found that SAM 2 is less effective on low contrast images. Accordingly, we continue with the original SAM model.

\subsubsection{Automatic mask generator (SAM-A)}
\label{s_sam-a}

SAM-A autonomously segments an image by selecting a set of sample points from a fixed grid that spans the entire input image (see Fig.~\ref{fig:samAuto_flow}a for a visual representation of the grid layout). For each selected point, the model generates one or more masks, where pixels corresponding to identified objects were labeled as feature pixels (value of one), and pixels representing the background were labeled as non-feature pixels (value of zero). SAM processes the masks to remove redundancy by identifying and eliminating overlaps or duplicates. It then filters the remaining masks to ensure quality, retaining only those with high predicted Intersection-over-Union (IoU) scores, indicating the model’s confidence in mask accuracy, and high stability scores, reflecting robustness to minor perturbations \citep{Kirillov2023segment}. The final output is a list of filtered masks representing all distinct elements in the image that SAM successfully segments (Fig.~\ref{fig:samAuto_flow}b).

SAM cannot differentiate between the various segmented elements and, hence, segmentation includes both ice floes and open water fractions. The latter is excluded by applying the convexity coefficient \citep{Sklansky_1970}  as a filtering criterion. The complexity coefficient quantifies how closely an object’s boundary resembles that of a convex shape. Ice floes, with regular boundaries that can often be approximated by ellipses or circles \citep{alberello2019brief}, typically exhibit convexity values close to 1. In contrast, open water regions, with more irregular boundaries, have lower convexity values. A cut-off value of 0.8 is set, such that masks with a convexity less than or equal to 0.8 are excluded \citep[cf.][]{wang2024integrating}. This threshold represents the minimum value that reliably removes water regions while preserving genuine floe masks across all images. The remaining masks (floes) are analyzed to identify those with anomalous shapes or open boundaries at the edges of the image using eccentricity and circularity metrics (see \S\ref{s_gvf}). Masks exhibiting these irregularities are also eliminated because they are either inconsistent with typical floe geometry or incompletely segmented due to edge effects.

Closer examination of the output reveals that some masks represent the same floes multiple times (see the first two layers in Fig.~\ref{fig:samAuto_flow}b), indicating that the model’s de-duplication process is not fully effective. This is common with adjacent or overlapping floes that have poorly defined boundaries. The redundant masks are then autonomously consolidated into a single unified mask \citep[cf.][]{shankar2023semantic}, through a pixel-wise union operation, where each pixel is marked as a feature pixel in the final binary mask if labeled as such in any initial mask. This process removes redundancy, ensuring that multiple occurrences of the same feature pixels across different masks are counted only once in the final mask (see Fig.~\ref{fig:samAuto_flow}c).

Geometrical properties of floes and sea ice concentration are extracted from the final unified binary mask. The runtime for segmentation and floe analysis is approximately 0.8\,s per floe regardless of floe morphology.

\subsubsection{Prompt-based (SAM-P)}
\label{sec_promptbased}

In prompt-based mode, segmentation is guided by user-defined seeds, which are-derived from the distance transform matrix introduced in \S\ref{s_gvf} (Fig.~\ref{fig:gvf_outputs}b). These seeds pre-identify the floes, and, for each one, SAM-P generates a corresponding mask (Fig.~\ref{fig:samAuto_flow}d). In regions with poorly defined boundaries, the assignment of seeds to floes can be ambiguous, and the resulting mask does not always isolate the intended floe accurately. In such cases, the mask includes unwanted areas, such as discontinuous water regions or parts of adjacent floes (see the first two layers in Fig.~\ref{fig:samAuto_flow}e). Shape-based filters for circularity, eccentricity, and convexity, with user-defined thresholds (see \S\ref{s_gvf} and \S\ref{s_sam-a}), are applied to remove clusters of feature pixels that do not match the typical shape of ice floes. A union operation is also used to merge clusters representing parts of adjacent floes with the correct floe. This operation combined all the masks in the output list into a single unified mask, automatically consolidating the segmented portions of floes into their appropriate masks (Fig.~\ref{fig:samAuto_flow}f). From this final binary mask, the properties of each individual ice floe and sea ice concentration are extracted.

The total runtime is around 0.3\,s per floe, including the generation of seeds from the distance transform matrix, making SAM-P more than twice as fast as SAM-A. This is because the prompt-based mode bypasses SAM’s internal point selection process, speeding up segmentation.

\subsection{A combination of SAM and GVF (SAM-GVF)}

\begin{figure}
\centering\includegraphics[width=0.9\textwidth]{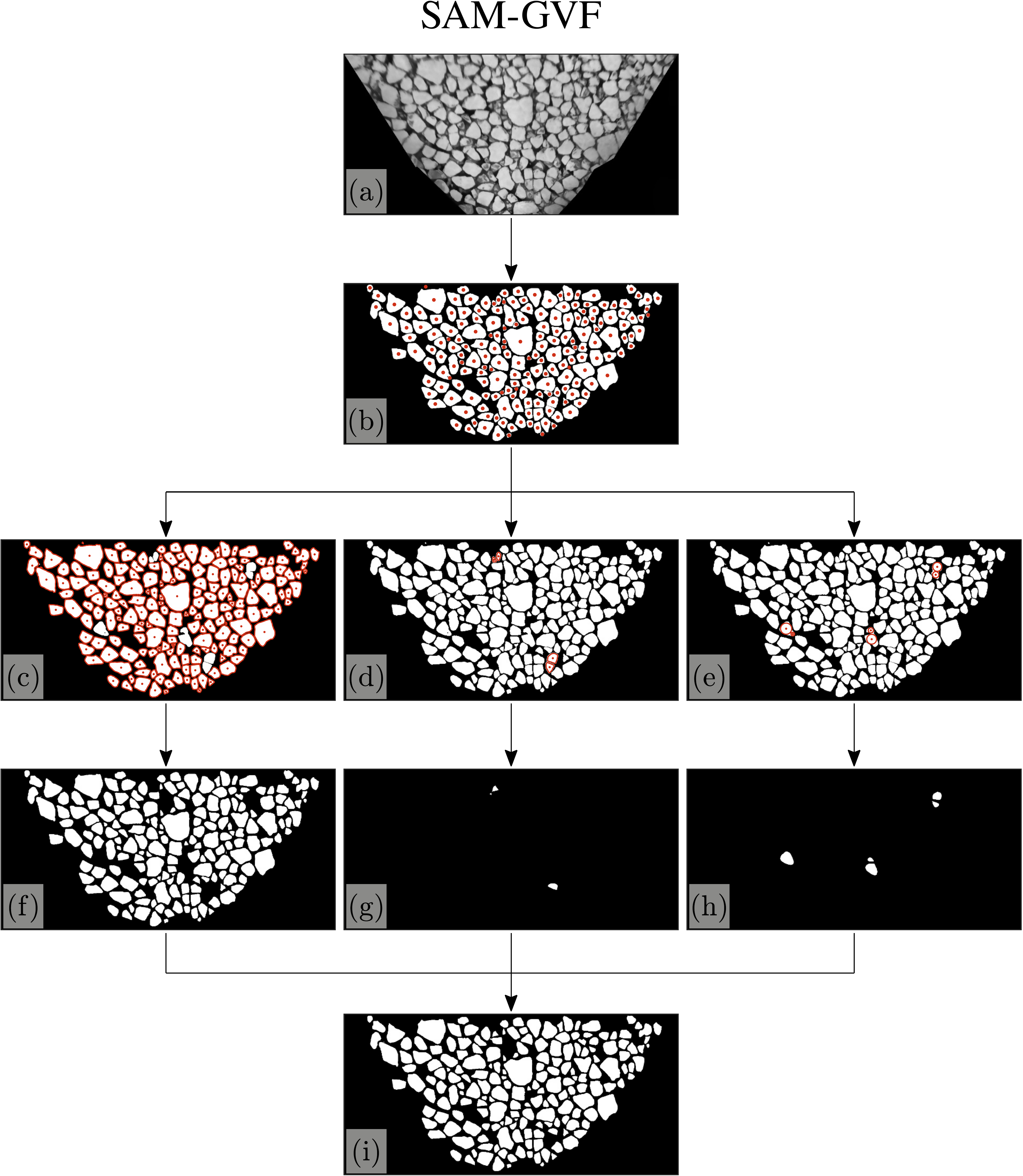}
\caption{SAM-GVF hybrid segmentation approach: (a)~enhanced sea ice image; (b)~unified binary mask from SAM-P showing seeds (red dots) of floes; (c)~perimeters (red lines) of floes correctly segmented by SAM-P; (d)~initial contours (red lines) of floes containing a single seed after morphological erosion; (e)~circular initial contours (red lines) for remaining unused seed; (f)~floes identified from SAM-P segmentation (as shown in Fig.~\ref{fig:samgvf_flow}c); (g)~floes detected by GVF initialized from initial contours in Fig.~\ref{fig:samgvf_flow}d; (h)~floes detected by GVF initialized from circular contours; (i) final binary segmentation with all detected floes.}
\label{fig:samgvf_flow}
\end{figure}

Our analysis indicates that the GVF algorithm efficiently delineates floes but encounters difficulties with varied gradients in gray-scale intensity due to its reliance on visual characteristics. SAM, on the other hand, can process a wide range of image variations due to its extended training, but it lacks shape integration, making it difficult to segment adjacent floes with weak edges. We propose combining SAM with GVF to compensate for shortcomings of the former by incorporating shape details, an approach conceptually similar to those suggested previously \citep[e.g.,][]{chen2019learning,hoangnganle2020active}.

SAM-P is applied to generate an initial binary mask (Fig.~\ref{fig:samgvf_flow}b), from which 90\% of the floes are detected, on average (Fig.~\ref{fig:samgvf_flow}b and f). Adjacent floes with blurred or unclear edges are captured by SAM as objects but were not segmented due to the lack of distinct boundaries. The GVF model is applied to analyze these floes. Objects connected by only a few pixels are separated using morphological erosion, resulting in newly delineated edges that serve as initial contours for the GVF model (Fig.~\ref{fig:samgvf_flow}d). For objects sharing significant edge portions, circles are used as initial contours (Fig.~\ref{fig:samgvf_flow}e). These contours are then expanded by the GVF algorithm to trace the true boundaries of each object (Fig.~\ref{fig:samgvf_flow}g-h). Anomalies, such as spurious features with nearly perfect circles or elongated line segments, are removed using circularity and eccentricity coefficients.

The final SAM-GVF  output(Fig.~\ref{fig:samgvf_flow}i) is processed to extract floe properties and sea ice concentration. Segmentation and analysis had a runtime ranging from approximately 4\,s to 6\,s per floe.

\section{Comparison against manual segmentation}
\label{sec:results}

\subsection{Benchmark dataset}

A benchmark is created by manually outlining a subset of twenty images from the original dataset (see \S\ref{benchdata}). These images were sampled at random and then reviewed to ensure they excluded scenes dominated by open water or consolidated pack ice. The final set span a wide range of visual conditions and varying levels of sea ice clarity. Segmentation of individual floes is performed using image editing software (Pixelmator, though other tools could also be used). While this manual process provides high-quality reference data, it is inherently subject to a degree of subjectivity, especially in complex ice conditions where floe boundaries may be ambiguous. However, when performed by trained sea ice observers---in this study, three experienced observers who cross-checked a few sample images---the segmentation results were generally consistent and reproducible. Furthermore, to ensure temporal consistency in challenging cases, observers also referred to video recordings captured by the camera, cross-checking ambiguous floe boundaries against subsequent frames.

Reference data for the number and size of floes, as well as ice concentration, are subsequently derived. A total of 2,874 floes are identified across the twenty benchmark images, with an average diameter of 3.3 m and an average ice concentration of 45\%, providing a solid foundation for benchmarking \citep{Arbelaez2011,MartinFTM01}. The same images are segmented using the automated methods (GVF, SAM-A, SAM-P, and the combined SAM-GVF). Segmentation performance of these methods is summarized in Table~\ref{tableX}, categorized by three types of errors: missing floes, where floes identified in the manual segmentation are not detected by the automated methods; misinterpreted floes, which are either partially segmented or merged with adjacent floes; and ``hallucinations'', referring to false objects where patterns or textures are mistakenly identified as floes.

\subsection{Overall segmentation performance}

Compared with manual segmentation of the 20-image dataset, the GVF method exhibits the highest number of missed floes (1,030), accounting for approximately 35\% of the total. SAM-P, in contrast, misses the fewest floes ($\approx$20\%), while SAM-A and SAM-GVF each miss slightly more than 25\%. Across all methods, missed floes have an average diameter of around 2\,m, covering between 2\% and 5\% of the benchmark image area. The exception is SAM-A, for which missed floes are slightly smaller (1.7\,m), suggesting that SAM, when segmenting autonomously without seed guidance, preferentially detects larger floes while overlooking smaller ones.

\begin{table}[htbp]
\centering
\caption{Segmentation results for the entire benchmark dataset (20 images) using the four methods. For each method, the table reports the number of floes (Fl.), mean diameter ($\overline{\mathrm{D}}$) and ice coverage (IC) associated with the three segmentation errors (missing floes, misinterpreted floes, hallucinations).}
\label{tableX}
\renewcommand{\arraystretch}{1.3}  
\begin{tabular}{l|ccc|ccc|ccc}
\toprule
\multicolumn{10}{c}{\textbf{Benchmark dataset}}\\
\midrule
\multirow{2}{*}[-0.25em]{Method}
 & \multicolumn{3}{c|}{Missing}
 & \multicolumn{3}{c|}{Misinterpreted}
 & \multicolumn{3}{c}{Hallucinations}\\
\cmidrule(lr){2-10}
 & Fl. & $\overline{\mathrm{D}}$ & IC
 & Fl. & $\overline{\mathrm{D}}$ & IC
 & Fl. & $\overline{\mathrm{D}}$ & IC\\
\midrule
GVF
 & 1030 & 2.09\,m & 5\%
 & 423  & 3.54\,m & 7\%
 & 198  & 1.79\,m & 1\%\\
SAM-A
 & 767 & 1.66\,m & 2\%
 & 241 & 5.37\,m & 8\%
 & 97  & 3.40\,m & 1\%\\
SAM-P
 & 579 & 2.00\,m & 2\%
 & 229 & 4.45\,m & 5\%
 & 654 & 1.13\,m & 1\%\\
SAM-GVF
 & 707 & 1.97\,m & 3\%
 & 167 & 3.99\,m & 3\%
 & 384 & 1.36\,m & 1\%\\
\bottomrule
\end{tabular}
\end{table}

Misinterpretations range from 6\% to 15\% of the total floes, with SAM-GVF performing best (167 misinterpretations) and GVF performing worst (423 misinterpretations). These misinterpreted floes typically have average diameters ranging between 3.5\,m and 5.4\,m, larger than those of the missed floes, indicating that these errors mostly result from merging adjacent floes into larger objects.

Hallucinations vary significantly among methods but generally affect only around 1\% of the total image area, which can be considered marginal in terms of their impact on overall segmentation accuracy. The average diameter of hallucinated floes is typically below 2\,m, except for SAM-A, which produces hallucinations averaging 3.4\,m. This further confirms that SAM-A, even when incorrectly segmenting, tends to detect relatively larger objects. The marked variability in hallucinations, particularly notable for SAM-P and SAM-GVF, primarily arises from image ``22-07-24 02-52-21'' (Fig.~\ref{fig:dataset2}), which alone accounts for more than 35\% of their total hallucinations. In this image, snowfall is misclassified as small floes, substantially increasing the hallucination count. For such cases, none of the methods may be optimal, and alternative, more stringent pre-processing strategies, such as cropping images to retain only areas of higher clarity, should be considered to effectively reduce noise and improve segmentation accuracy.

Overall, these results highlight notable variability in segmentation performance across different methods. A detailed analysis of specific cases is presented in the following section to further clarify how segmentation methods respond to the inherent heterogeneity in sea ice and image clarity.

\subsection{Segmentation performance on sample images}
\label{sampl2img}

Four sample images representing contrasting ice conditions are used to evaluate the segmentation methods against manual segmentation, each highlighting specific challenges. Two cases primarily focus on floe boundaries: one features distinct floes with well-defined edges, while the other exhibits diffuse boundaries due to closely packed or interconnected floes and interstitial ice. The remaining two cases examine floe shape: one with uniformly round circular floes, and another characterized by angular square-shaped floes. This comparison illustrates how varying sea ice conditions influence segmentation outcomes and provides deeper insight into model performances (within their respective constraints).

Figure~\ref{fig:differences1} and Table~\ref{table1} present results for the image with distinct floe edges (``22-07-23 12-06-06'' in Fig.~\ref{fig:dataset2}). This image has sharp contrast, uniform lighting, and low noise level, reducing ambiguities. Figure~\ref{fig:differences2} and Table~\ref{table2} report results for the more challenging diffuse boundaries case (``22-07-23 09-58-14'' in Fig.~\ref{fig:dataset2}), featuring reduced contrast, subtle lighting variations, and interstitial ice. These factors complicate the detection of floe boundaries, making the automated segmentation more prone to merging distinct floes, overlooking small or faint floes, and introducing false floes where weak edges or texture gradients are mistaken for true boundaries.

Figure~\ref{fig:differences3} and Table~\ref{table3} illustrate results for the round circular floes image (``22-07-24 11-17-13'' in Fig.~\ref{fig:dataset2}), characterized by medium-sized floes with uniform shapes and rounded corners. Figure~\ref{fig:differences4} and Table~\ref{table4} detail results for the angular square floes image (``22-07-21 12-48-26'' in Fig.~\ref{fig:dataset1}), which contains generally larger, irregularly shaped floes alongside smaller floes with complex morphology. These angular floes, typically formed by merging several smaller pancakes, have intricate internal textures that present additional segmentation challenges. In both cases, floes are well separated, isolating shape as the primary factor influencing segmentation performance.

\subsubsection{Distinct ice floes with defined edges}

\begin{figure}
\centering\includegraphics[width=0.8\textwidth]{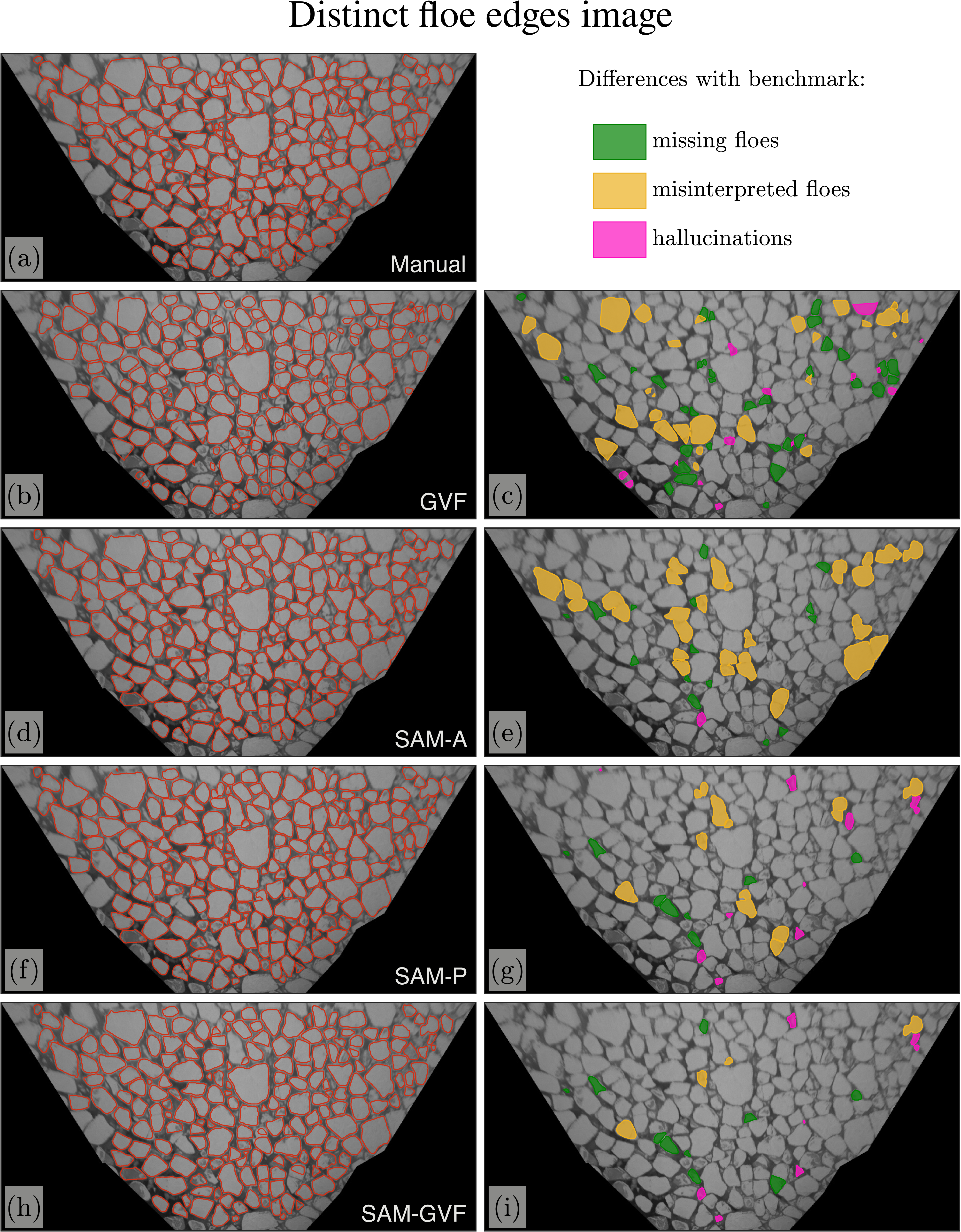}
\caption{Visual comparison between manual and automated segmentation methods for an image with distinct ice floes and defined edges from the benchmark dataset (``22-07-23 12-06-06'' in Fig.~\ref{fig:dataset2}). The left column displays the original image overlaid with segmented floe contours (red), showing the manual benchmark~(a) and automated methods: GVF~(b), SAM-A~(d), SAM-P~(f), and combined SAM-GVF~(h). The right column illustrates segmentation differences relative to the manual benchmark (c, e, g, i), highlighting missing floes (green), misinterpreted floes (yellow), and hallucinations (pink).}
\label{fig:differences1}
\end{figure}

The manual segmentation identifies 196 floes with an average diameter of 3.9\,m and an ice concentration of 57\%. The GVF method misses 35 floes, misinterprets 29, and produces 12 hallucinations (Fig.~\ref{fig:differences1}c), resulting in approximately 40\% of floes not being correctly identified. The missing floes have an average diameter of 2.4\,m, representing about 4\% of the total surface area. Most misinterpretations are due to partial segmentation of relatively large floes (average diameter of 3.3\,m), with few instances of adjacent floes being merged. The hallucinations consist of smaller floes, with an average diameter of 1.3\,m. Despite these errors, the GVF method appears to capture the general characteristics of the floe population. However, the relatively high number of missing floes and hallucinations introduces uncertainties in the ice concentration when compared to the manual benchmark.

\begin{table}[htbp]
\centering
\caption{Segmentation results for the distinct floe edges image (``22-07-23 12-06-06'') using the four methods. For each method, the table reports the number of floes (Fl.), mean diameter ($\overline{\mathrm{D}}$) and ice coverage (IC) associated with the three segmentation errors (missing floes, misinterpreted floes, hallucinations), along with the average computational segmentation time per floe.}
\label{table1}
\renewcommand{\arraystretch}{1.3}
\begin{tabular}{l|ccc|ccc|ccc|c}
\toprule
\multicolumn{11}{c}{\textbf{Distinct floe edges image}}\\
\midrule
\multirow{2}{*}[-0.25em]{Method}
 & \multicolumn{3}{c|}{Missing}
 & \multicolumn{3}{c|}{Misinterpreted}
 & \multicolumn{3}{c|}{Hallucinations}
 & \multirow{2}{*}[-0.25em]{\shortstack{Average\\CPU time\\per floe}}\\
\cmidrule(lr){2-10}
 & Fl. & $\overline{\mathrm{D}}$ & IC
 & Fl. & $\overline{\mathrm{D}}$ & IC
 & Fl. & $\overline{\mathrm{D}}$ & IC\\ 
\midrule
GVF
 & 35 & 2.39\,m & 4\%
 & 29 & 3.29\,m & 7\%
 & 12 & 1.31\,m & \textless{}1\%
 & 70\,s\\
SAM-A
 & 15 & 1.95\,m & 1\%
 & 19 & 5.29\,m & 10\%
 & 1  & 2.61\,m & \textless{}1\%
 & 0.8\,s\\
SAM-P
 & 8 & 2.49\,m & 1\%
 & 9 & 4.25\,m & 3\%
 & 10 & 1.84\,m & 1\%
 & 0.3\,s\\
SAM-GVF
 & 10 & 2.63\,m & 1\%
 & 4  & 3.44\,m & 1\%
 & 7  & 1.77\,m & \textless{}1\%
 & 4\,s\\
\bottomrule
\end{tabular}
\end{table}

SAM-A misses 15 floes, misinterprets 19, and generates one hallucination (approximately 18\% of the floes; Fig.~\ref{fig:differences1}e). The missed floes have an average diameter of 2\,m, covering about 1\% of the image area, resulting in a more consistent ice concentration estimate than GVF for this specific image. Most misinterpretations result from adjacent floes merging, producing larger floes with an average diameter of 5.3\,m and, hence, introducing a bias toward larger floe sizes. The single hallucination covers less than 1\% of the image area, making its impact on overall sea ice characteristics negligible.

SAM-P misses only eight floes, and, thus, is the best performing method for capturing floes (Fig.~\ref{fig:differences1}g). The missed floes have an average diameter of 2.5\,m and cover 1\% of the image area, resulting in an accurate estimate of ice concentration relative to the benchmark. SAM-P misinterprets nine floes and generates ten hallucinations, bringing the total proportion of incorrectly identified floes to less than 15\%. The misinterpreted floes have an average diameter of 4.3\,m and are mainly due to merged floes, similar to SAM-A. The number of hallucinations is higher than in SAM-A and is attributed to the model’s susceptibility to complex textures when using prompts to identify uncertain regions \citep{fan2023stable}. These hallucinations have an average diameter of 1.8\,m and cover 1\% of the image area, having minimal impact on the overall floe population and sea ice concentration.

SAM-GVF misses ten floes, misinterprets four, and produces seven hallucinations ($\approx$11\% of floes). The missing floes have an average diameter of 2.6\,m, covering around 1\% of the surface area, similar to SAM-A and SAM-P. Misinterpreted floes have an average diameter of 3.4\,m and include both partially segmented and merged floes. Hallucinations have an average diameter of 1.8\,m and align with those seen in SAM-P.

Overall, SAM-GVF has the lowest percentage of segmentation errors, with uncertainties affecting $<$3\% of the image area. SAM-P achieves similar performance ($\approx$5\%), while SAM-A and GVF exhibit the highest uncertainty, affecting 11\% and 12\% of the surface area, respectively.

\subsubsection{Diffuse sea ice boundaries}

\begin{figure}
\centering\includegraphics[width=0.8\textwidth]{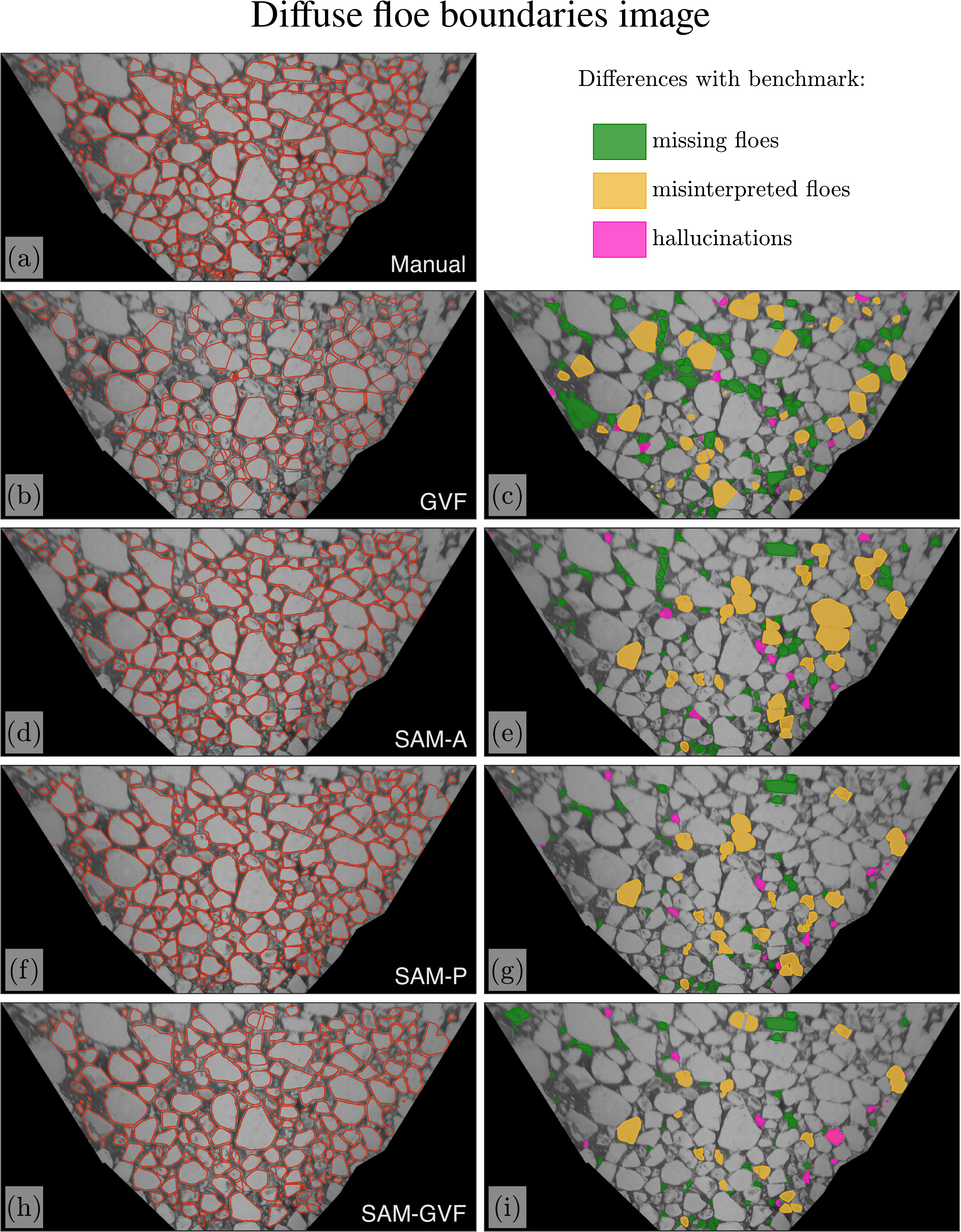}
\caption{Visual comparison between manual and automated segmentation methods for an image with diffuse sea ice boundaries from the benchmark dataset (``22-07-23 09-58-14'' in Fig.~\ref{fig:dataset2}). The left column displays the original image overlaid with segmented floe contours (red), showing the manual benchmark~(a) and automated methods: GVF~(b), SAM-A~(d), SAM-P~(f), and combined SAM-GVF~(h). The right column illustrates segmentation differences relative to the manual benchmark (c, e, g, i), highlighting missing floes (green), misinterpreted floes (yellow), and hallucinations (pink).}
\label{fig:differences2}
\end{figure}

As expected, segmentation performance declines under more challenging conditions. For the sample image with diffuse boundaries, the manual segmentation detects 250 floes with an average diameter of 3\,m and an ice concentration of 55\%. The performance of GVF significantly degrades---it misses 104 floes, nearly three times the number observed in the distinct edges image. Misinterpretations also increase, with 40 floes affected, while the 17 hallucinations is similar to the distinct edges case (Fig.~\ref{fig:differences2}c). In total, $\approx$65\% of floes are not interpreted correctly, where the missed floes have an average diameter of 1.8\,m and cover approximately 7\% of the image area. The misinterpretations have an average diameter of 3\,m and result mostly from partial segmentation of larger floes. The hallucinations have an average diameter of 1.1\,m and do not substantially affect the overall sea ice characteristics. However, the increased number of missed and misinterpreted floes reduces the reliability of ice concentration estimates in the diffuse boundaries image.

\begin{table}[htbp]
\centering
\caption{Segmentation results for the diffuse floe boundaries image (``22-07-23 09-58-14'') using the four methods. For each method, the table reports the number of floes (Fl.), mean diameter ($\overline{\mathrm{D}}$) and ice coverage (IC) associated with the three segmentation errors (missing floes, misinterpreted floes, hallucinations), along with the average computational segmentation time per floe.}
\label{table2}
\renewcommand{\arraystretch}{1.3}
\begin{tabular}{l|ccc|ccc|ccc|c}
\toprule
\multicolumn{11}{c}{\textbf{Diffuse floe boundaries image}}\\
\midrule
\multirow{2}{*}[-0.25em]{Method}
 & \multicolumn{3}{c|}{Missing}
 & \multicolumn{3}{c|}{Misinterpreted}
 & \multicolumn{3}{c|}{Hallucinations}
 & \multirow{2}{*}[-0.25em]{\shortstack{Average\\CPU time\\per floe}}\\
\cmidrule(lr){2-10}
 & Fl. & $\overline{\mathrm{D}}$ & IC
 & Fl. & $\overline{\mathrm{D}}$ & IC
 & Fl. & $\overline{\mathrm{D}}$ & IC\\ 
\midrule
GVF
 & 104 & 1.77\,m & 7\%
 & 40 & 2.97\,m & 8\%
 & 17 & 1.10\,m & \textless{}1\%
 & 75\,s\\
SAM-A
 & 81 & 1.30\,m & 3\%
 & 20 & 4.49\,m & 8\%
 & 10 & 1.86\,m & 1\%
 & 0.8\,s\\
SAM-P
 & 53 & 1.18\,m & 2\%
 & 22 & 3.25\,m & 5\%
 & 16 & 1.19\,m & 1\%
 & 0.3\,s\\
SAM-GVF
 & 58 & 1.24\,m & 2\%
 & 20 & 2.84\,m & 3\%
 & 11 & 1.24\,m & \textless{}1\%
 & 5\,s\\
\bottomrule
\end{tabular}
\end{table}

SAM-A is more robust for the diffuse boundaries image in comparison to GVF, although it shows a significant decline in performance. Approximately 81 floes are missed, which is over five times more than for the distinct edges image. Misinterpretations remain relatively stable at twenty floes, while hallucinations increase to ten, although this is still fewer than GVF (Fig.~\ref{fig:differences2}e). Overall, around 45\% of floes are not captured correctly. The missed floes have an average diameter of 1.3\,m, covering about 3\% of the image area. While SAM-A misses more floes in the diffuse boundaries image compared to the distinct edges image, it still retains the capacity to provide an accurate estimate of ice concentration. Misinterpretations have an average floe diameter of 4.5\,m, and primarily arise from the merging of adjacent floes, continuing SAM-A’s tendency to bias floe size distribution toward larger floes. Despite the increased hallucinations, their overall impact on sea ice characteristics remains limited.

SAM-P also experiences an increase in segmentation errors, missing 53 floes, which is greater than six times more than in the distinct edges sample. The missed floes are small, with an average diameter of 1.2\,m and covering only 2\% of the total area. Misinterpretations double to 22 instances, involving smaller floes (average diameter of 3.3\,m) than those in the distinct edges image. These misinterpretations primarily result from partial segmentation of floes featuring both snow-covered and bare ice surfaces, which subsequently affects the estimated floe size. Hallucinations increase slightly to 16, likely due to the model’s sensitivity to complex textures when using prompts to refine uncertain regions. Approximately 36\% of floes are not interpreted correctly.

SAM-GVF shows moderate performance degradation, missing 58 floes, which is a substantial increase (almost sixfold) compared to the distinct edges image. Misinterpretations remain relatively low, with 20 floes affected, while hallucinations increase slightly to 11, which is fewer than SAM-P and GVF (Fig.~\ref{fig:differences2}i). Similar to SAM-P, $\approx$36\% of the floes are not captured correctly. The missed floes have an average diameter of 1.2\,m, covering about 2\% of the image area. Misinterpretations, with an average floe diameter of 2.8\,m, mainly result from partial segmentation of larger floes and those with mixed snow and bare ice surfaces, which slightly distorts the floe size distribution. Hallucinations remain small, with an average diameter of 1.2\,m, and do not significantly impact the overall sea ice characteristics.

Of the various methods, SAM-P provides the best overall balance of accuracy and computational efficiency across both the distinct edges and diffuse boundaries images. SAM-P has the fewest missed floes and maintains stable, though slightly higher, rates of misinterpretations and hallucinations compared to SAM-GVF. While SAM-GVF performs marginally better in terms of misinterpretation and hallucination metrics, these improvements come at a substantial cost in computational time ($\approx$15 times slower per floe). SAM-A and GVF show significant deterioration, particularly with a substantial increase in missed floes. Additionally, computational time per floe increases for GVF (and consequently for SAM-GVF) in the diffuse boundaries image, whereas it remains steady in SAM.

\subsubsection{Round circular ice floes}

\begin{figure}
\centering\includegraphics[width=0.8\textwidth]{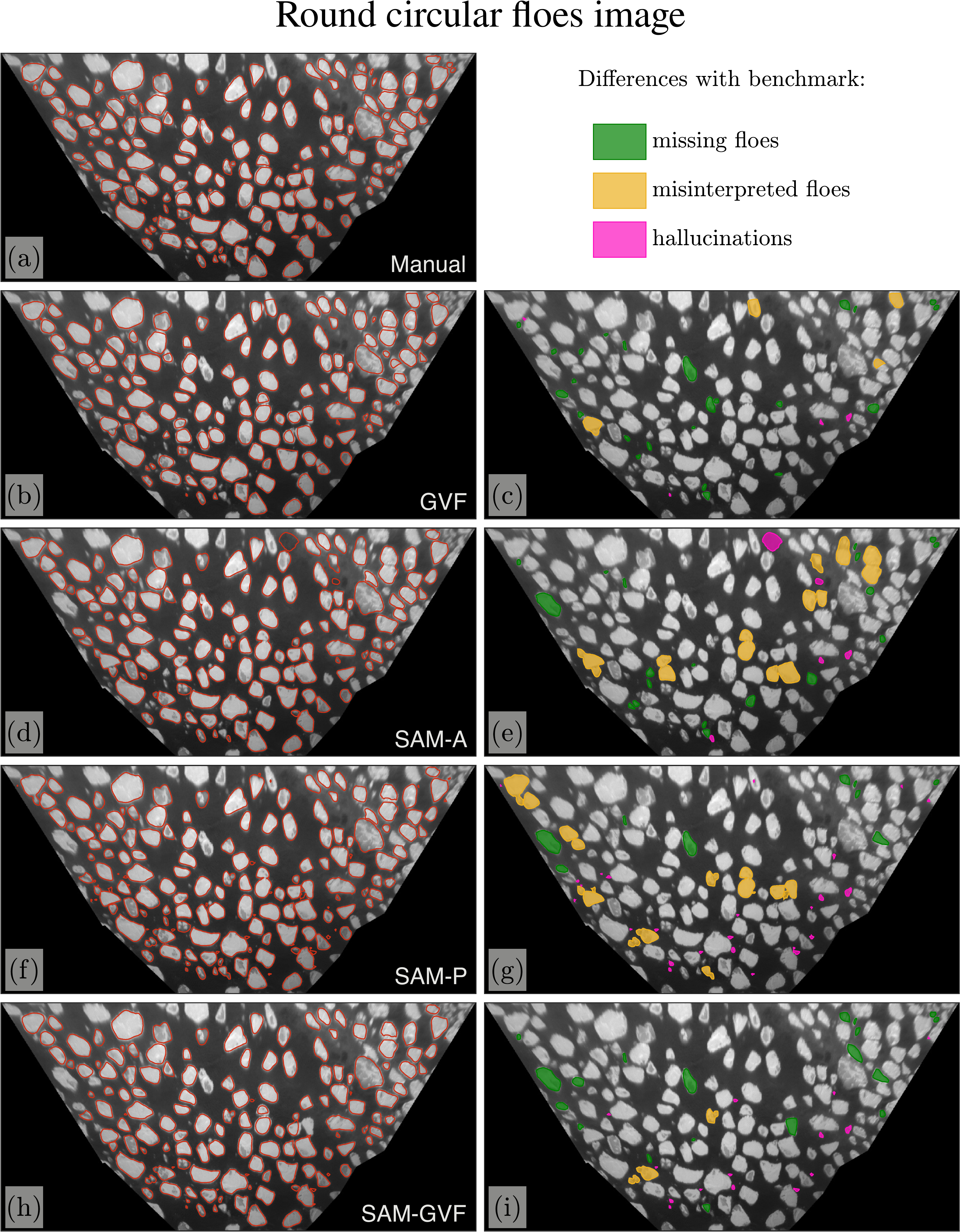}
\caption{Visual comparison between manual and automated segmentation methods for an image with round circular ice floes from the benchmark dataset (``22-07-24 11-17-13'' in Fig.~\ref{fig:dataset2}). The left column displays the original image overlaid with segmented floe contours (red), showing the manual benchmark~(a) and automated methods: GVF~(b), SAM-A~(d), SAM-P~(f), and combined SAM-GVF~(h). The right column illustrates segmentation differences relative to the manual benchmark (c, e, g, i), highlighting missing floes (green), misinterpreted floes (yellow), and hallucinations (pink).}
\label{fig:differences3}
\end{figure}

The manual segmentation identifies a total of 145 floes, averaging 3.31\,m in diameter with an ice concentration of 31\%. Using the GVF method, 24 floes are missed, four are misinterpreted, and four hallucinations are produced, collectively accounting for approximately 22\% of floes not correctly identified (Fig.~\ref{fig:differences3}c). Missed floes have an average diameter of 1.5\,m, misinterpretations average 3.5\,m, and hallucinations average 0.8\,m. Together, these errors account for less than 3\% of the total surface area. Most misinterpretations arise from partial segmentation of larger floes, while hallucinations and misses involve smaller floes, indicating that GVF has difficulty detecting fine-scale features. Nevertheless, these errors minimally impact the final results, particularly when averaging floe diameters across the entire image.

\begin{table}[htbp]
\centering
\caption{Segmentation results for the round circular floes (``22-07-24 11-17-13'') image using the four methods. For each method, the table reports the number of floes (Fl.), mean diameter ($\overline{\mathrm{D}}$) and ice coverage (IC) associated with the three segmentation errors (missing floes, misinterpreted floes, hallucinations), along with the average computational segmentation time per floe.}
\label{table3}
\renewcommand{\arraystretch}{1.3}
\begin{tabular}{l|ccc|ccc|ccc|c}
\toprule
\multicolumn{11}{c}{\textbf{Round circular floes image}}\\
\midrule
\multirow{2}{*}[-0.25em]{Method}
 & \multicolumn{3}{c|}{Missing}
 & \multicolumn{3}{c|}{Misinterpreted}
 & \multicolumn{3}{c|}{Hallucinations}
 & \multirow{2}{*}[-0.25em]{\shortstack{Average\\CPU time\\per floe}}\\
\cmidrule(lr){2-10}
 & Fl. & $\overline{\mathrm{D}}$ & IC
 & Fl. & $\overline{\mathrm{D}}$ & IC
 & Fl. & $\overline{\mathrm{D}}$ & IC\\ 
\midrule
GVF
 & 24 & 1.51\,m & 1\%
 & 4  & 3.45\,m & 1\%
 & 4  & 0.80\,m & \textless{}1\%
 & 35\,s\\
SAM-A
 & 19 & 1.66\,m & 1\%
 & 8  & 5.31\,m & 4\%
 & 5  & 2.06\,m & \textless{}1\%
 & 0.8\,s\\
SAM-P
 & 9  & 2.44\,m & 1\%
 & 9  & 4.41\,m & 3\%
 & 25 & 0.69\,m & \textless{}1\%
 & 0.3\,s\\
SAM-GVF
 & 19 & 2.13\,m & 2\%
 & 3  & 3.05\,m & \textless{}1\%
 & 14 & 0.73\,m & \textless{}1\%
 & 6\,s\\
\bottomrule
\end{tabular}
\end{table}

SAM-A demonstrates performance similar to GVF in terms of error quantity, missing 19 floes, misinterpreting eight, and generating five hallucinations (also totaling approximately 22\%). The missing floes are slightly larger (1.6\,m average diameter) and cover a similar area to GVF. Misinterpretations exhibit a larger average diameter (5.3\,m), primarily due to the merging of adjacent floes, biasing the size distribution toward larger values. However, their limited number ensures a minimal impact on the overall results. Hallucinations average 2.1\,m in diameter, arising mainly from darker oceanic features mistaken as floes (Fig.~\ref{fig:differences3}e), yet they affect less than 1\% of the total area.

SAM-P misses the fewest floes (only nine), with nine misinterpretations and notably higher hallucinations (25), totaling approximately 43\% of incorrectly identified floes. Misinterpretations have a slightly smaller average diameter (4.4\,m) compared to SAM-A, as the model correctly separates some previously merged floes (Fig.~\ref{fig:differences3}g). Missing floes tend to be larger (2.4\,m), frequently occurring when one floe in a merged pair is correctly identified while the other is missed entirely. Hallucinations, though numerous, have a mean diameter of 0.7\,m, primarily arising from image noise near larger floes, and cover less than 1\% of the area. In total, all errors combined affect less than 4\% of the image area, confirming SAM-P's strong ability to accurately estimate floe characteristics and ice concentration.

SAM-GVF misses more floes than SAM-P (19 floes, similar to SAM-A and GVF), but has fewer misinterpretations (only three) and a small average misinterpretation diameter (3.05\,m). Hallucinations are similar to SAM-P, averaging 0.7\,m and covering less than 1\% of the area. Combined, SAM-GVF's errors represent 25\% of the floes, consistent with SAM-A and GVF, but with minimal impact on ice concentration estimates.

Overall, the round circular floes image represents optimal conditions, featuring relatively few, well-spaced floes with uniform shape and size. Consequently, all methods perform well, particularly in capturing average floe diameters. Under such optimal conditions, computational efficiency becomes especially important: SAM-P is notably the fastest method (0.3\,s per floe), SAM-A takes slightly more than twice as long, SAM-GVF is roughly twenty times slower, and GVF is the slowest, requiring approximately 35\,s per floe.

\subsubsection{Angular square ice floes}

\begin{figure}
\centering\includegraphics[width=0.8\textwidth]{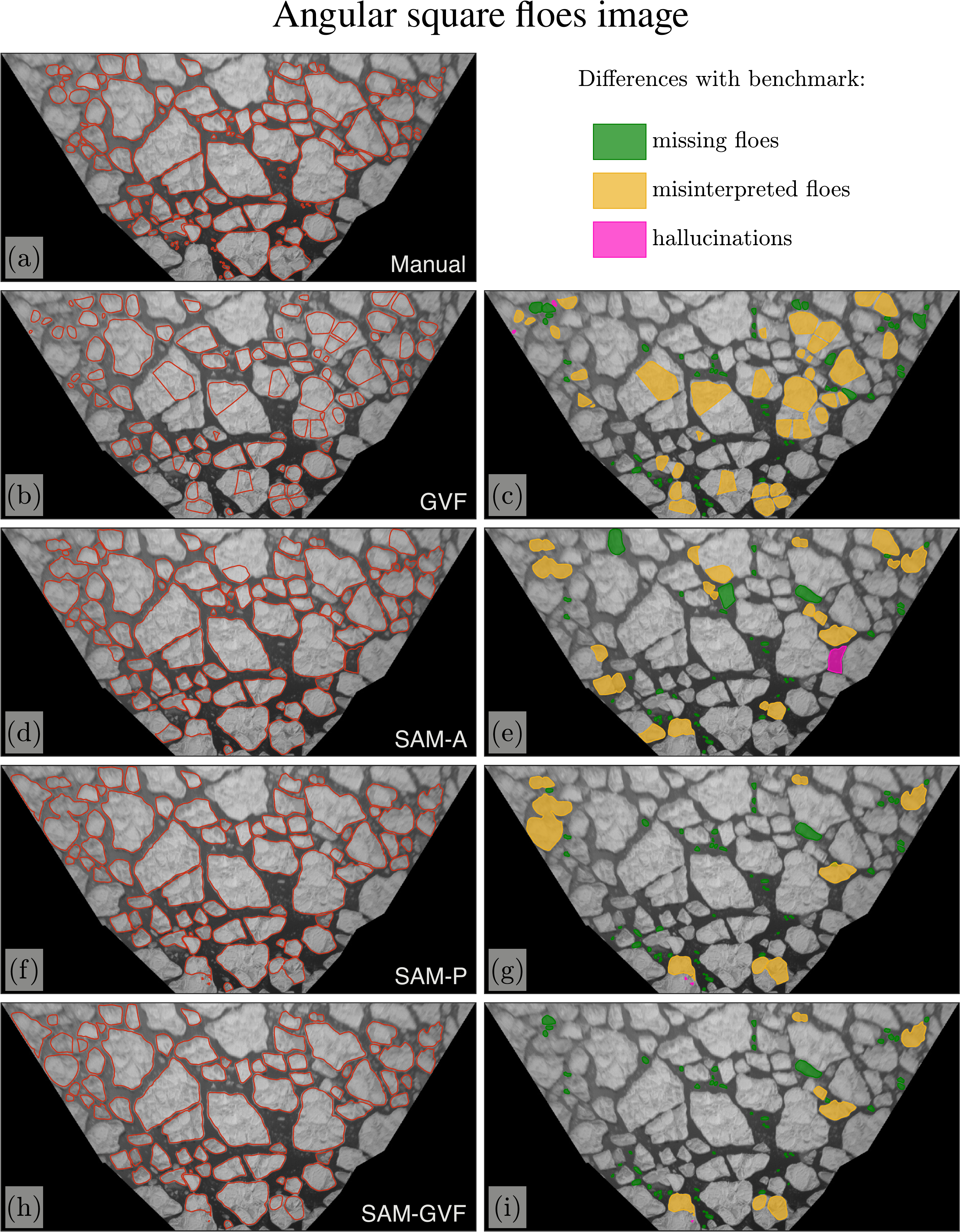}
\caption{Visual comparison between manual and automated segmentation methods for an image with angular square ice floes from the benchmark dataset (``22-07-21 12-48-26'' in Fig.~\ref{fig:dataset1}). The left column displays the original image overlaid with segmented floe contours (red), showing the manual benchmark~(a) and automated methods: GVF~(b), SAM-A~(d), SAM-P~(f), and combined SAM-GVF~(h). The right column illustrates segmentation differences relative to the manual benchmark (c, e, g, i), highlighting missing floes (green), misinterpreted floes (yellow), and hallucinations (pink).}
\label{fig:differences4}
\end{figure}

Segmentation performance slightly degrades in the angular square floes scenario due to the less regular and more heterogeneous floe shapes and sizes. For this sample image, manual segmentation identifies 127 floes with an average diameter of 3.5\,m and an ice concentration of 49\%. The GVF method misses 59 floes, more than double the number compared to the round circular floes case. The missed floes have an average diameter of 1.2\,m, affecting approximately 2\% of the total image area, confirming their limited impact on ice concentration but significant influence on the smaller size tail of the floe distribution. A substantial difference emerges in misinterpretations, totaling 38 (nearly tenfold higher), with an average diameter of 4\,m, impacting 12\% of the surface area. These errors predominantly result from partial segmentations of larger, irregularly shaped floes, indicating GVF's difficulty in handling large floes formed by merged pancakes (Fig.~\ref{fig:differences4}c). Hallucinations remain minimal, consistent with the previous observations. Overall, nearly 80\% of the floes are incorrectly identified.

\begin{table}[htbp]
\centering
\caption{Segmentation results for the angular square floes image (``22-07-21 12-48-26'') using the four methods. For each method, the table reports the number of floes (Fl.), mean diameter ($\overline{\mathrm{D}}$) and ice coverage (IC) associated with the three segmentation errors (missing floes, misinterpreted floes, hallucinations), along with the average computational segmentation time per floe.}
\label{table4}
\renewcommand{\arraystretch}{1.3}
\begin{tabular}{l|ccc|ccc|ccc|c}
\toprule
\multicolumn{11}{c}{\textbf{Angular square floes image}}\\
\midrule
\multirow{2}{*}[-0.25em]{Method}
 & \multicolumn{3}{c|}{Missing}
 & \multicolumn{3}{c|}{Misinterpreted}
 & \multicolumn{3}{c|}{Hallucinations}
 & \multirow{2}{*}[-0.25em]{\shortstack{Average\\CPU time\\per floe}}\\
\cmidrule(lr){2-10}
 & Fl. & $\overline{\mathrm{D}}$ & IC
 & Fl. & $\overline{\mathrm{D}}$ & IC
 & Fl. & $\overline{\mathrm{D}}$ & IC\\ 
\midrule
GVF
 & 59 & 1.20\,m & 2\%
 & 38 & 3.94\,m & 12\%
 & 2  & 0.97\,m & \textless{}1\%
 & 70\,s\\
SAM-A
 & 42 & 1.18\,m & 2\%
 & 17 & 4.59\,m & 6\%
 & 1  & 5.07\,m & \textless{}1\%
 & 0.8\,s\\
SAM-P
 & 52 & 1.08\,m & 1\%
 & 7  & 6.18\,m & 5\%
 & 2  & 0.50\,m & \textless{}1\%
 & 0.3\,s\\
SAM-GVF
 & 55 & 1.09\,m & 1\%
 & 7  & 4.53\,m & 3\%
 & 1  & 0.30\,m & \textless{}1\%
 & 6\,s\\
\bottomrule
\end{tabular}
\end{table}

SAM-A exhibits comparable missed floes metrics to GVF, with 42 misses, double the round circular floes scenario. The missed floes' average diameter is slightly smaller (1.2\,m), impacting about 2\% of the surface area. Misinterpretations also double to 17, averaging 4.6\,m in diameter and covering 6\% of the area. While SAM-A effectively identifies larger floes despite irregular shapes, segmentation remains imperfect due to pronounced textures causing partial segmentation and merged adjacent floes (Fig.~\ref{fig:differences4}e). Only one significant hallucination occurs, clearly identifiable as an erroneous large water gap classified as a floe.

SAM-P misses 52 floes, approximately five times more than in the round circular scenario, averaging 1.1\,m and covering 1\% of the image area. However, misinterpretations are fewer (only 7), with a larger average diameter (6.2\,m), covering 5\% of the area, demonstrating consistent segmentation capabilities regardless of floe shapes. Hallucinations are negligible in both number and impact.

SAM-GVF closely matches SAM-P performance, missing 55 floes (average diameter 1.1\,m) and producing only one small hallucination (diameter 0.30 m) with no impact on overall sea ice characteristics. Misinterpretations remain at seven, but have smaller diameters averaging less than 4.5\,m and cover only 3\% of the image area, indicating improved accuracy in separating merged floes compared to SAM-P.

Overall, GVF shows notably higher errors compared to the SAM-based methods. In terms of surface area affected by errors, SAM-GVF performs best at 4\%, followed by SAM-P (6\%), SAM-A (8\%), and GVF (14\%). Computationally, SAM-P remains highly efficient at 0.3\,s per floe, unaffected by floe complexity, whereas GVF’s computational time doubles to 70\,s per floe. SAM-GVF maintains consistent computational efficiency at approximately 6\,s per floe, benefiting from SAM’s speed.

\subsection{Floe count, size distribution, and sea ice concentration across the entire sample}

\begin{figure}
\centering\includegraphics[width=0.8\textwidth]{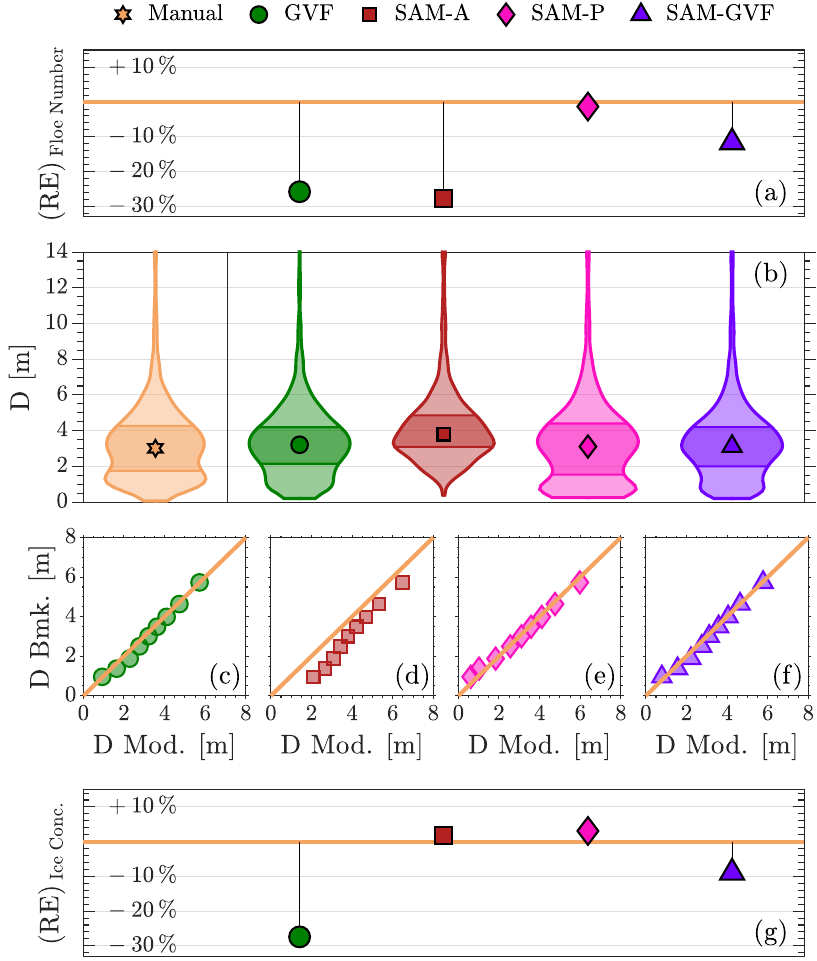}
\caption{Comparison of floe size statistics derived from segmentation methods GVF (green), SAM-A (red), SAM-P (pink), and SAM-GVF (purple) against the manual benchmark (orange). (a)~Relative error (RE) in the number of detected floes compared to the manual benchmark (horizontal orange line). b)~Violin plots of floe size distributions for each method, with the manual benchmark on the left. Squares inside violins represent medians; darker shaded areas indicate quartiles. (c,d,e,f)~QQ plots comparing deciles (10\% to 90\%) of floe size distributions from segmentation methods (D Mod.) against the manual benchmark (D Bmk.). The solid line indicates perfect alignment. (g)~Relative error (RE) in ice concentration compared to the manual benchmark (horizontal orange line).}
\label{fig:compar_stats}
\end{figure}

Data extracted from all 20 images in the sample are compared with the benchmark manual segmentation, as illustrated in Fig.~\ref{fig:compar_stats}. The relative error (RE) in estimating floe numbers is presented in Fig.~\ref{fig:compar_stats}a. Among the methods evaluated, SAM-P performs the best, slightly underestimating the number of floes by approximately 2\%, which highlights its capability to detect floes across images with varying clarity. In contrast, SAM-A and GVF are the least accurate, underestimating the manual count by approximately 28\% and 26\%, respectively. This reflects the difficulty in detecting floes in complex scenarios, particularly when dealing with unclear or irregular ice formations (cf.\ Figs.~\ref{fig:differences2} and \ref{fig:differences4}). For SAM-A, the automated initialization is a key limitation, as it prioritizes larger, more distinct floes while often overlooking smaller or poorly defined ones. The combined SAM-GVF approach mitigates some of the uncertainties in floe detection observed in both SAM-A and GVF, leading to a lower underestimation of the manual count (approximately 12\%). This aligns with its performance on the distinct edges image, suggesting that, like SAM-P, it remains relatively consistent across varying image conditions.

The floe size distribution is shown in Fig.~\ref{fig:compar_stats}b as violin plots, which display the overall distribution of data, including the median (marked inside the violins) and the first and third quartiles (represented by the darker shaded areas). The benchmark size distribution from manual segmentation exhibits a clear bimodal shape, with peaks around diameters of 1.5\,m and 3.5\,m. Approximately 50\% of the floes fall between 1.8\,m and 4.3\,m, with a median diameter of 3\,m. SAM-P aligns well with the benchmark, with the bimodal shape, median, and quartiles generally well captured. However, the distribution is slightly skewed toward smaller floes, as the first quartile underestimates the benchmark (1.5\,m instead of 1.8\,m), exaggerating the presence of smaller floes and reflecting the relatively large count of hallucinations (Fig.~\ref{fig:differences1}g). Both GVF and SAM-GVF methods also capture the overall shape of the distribution, including the median (3.2\,m and 3.1\,m, respectively) and quartiles. Minor discrepancies arise from the lower detection of smaller floes, resulting in a less pronounced bimodal shape compared to SAM-P. Conversely, SAM-A produces the least accurate distribution, missing the bimodal nature of the benchmark, resulting in a single peak around 3.3\,m. It overestimates both the median (3.8\,m) and quartiles, reflecting the limitations of SAM-A in detecting individual floes without seed-based prompts.

The comparison between benchmark and model distributions is further illustrated and analyzed in Figs.~\ref{fig:compar_stats}d-g, which present the deciles of the distributions as QQ plots. Despite differences in the overall bimodal shape, the deciles of SAM-P, SAM-GVF, and GVF align reasonably well with the benchmark (Figs.~\ref{fig:compar_stats}c,e,f). The lowest deciles (10--30\%) are slightly more pronounced, especially for SAM-P, suggesting that these methods detect a large number of small floes, attributed to spurious features or noise. However, the higher deciles align well with the benchmark. The agreement between the deciles can be further quantified using the root mean squared error (RMSE) of floe diameter, calculated as the square root of the mean squared differences between the observed (manual benchmark) and predicted (segmentation method) deciles of the floe size distribution. Using this metric, the RMSE is 0.20\,m for SAM-P, 0.17\,m for SAM-GVF, and 0.21\,m for GVF, which corresponds to approximately 3–4 times the smallest measurable unit in the images, given the 0.05\,m pixel-to-meter conversion in the orthorectification process. SAM-A (Fig.~\ref{fig:compar_stats}d) shows a more significant underestimation compared to the other models, especially in the lowest five deciles. The RMSE for SAM-A is 0.94\,m, reflecting its misalignment with the benchmark’s bimodality.

The relative error associated to sea ice concentration, relative to the benchmark, is shown in Fig.~\ref{fig:compar_stats}g. Both SAM-A and SAM-P yielded estimates within 3\% of the benchmark, with SAM-A performing slightly better. However, the accuracy of SAM-A is ambiguous, as it results from compensating factors, which are the detection of fewer floes (Fig.~\ref{fig:compar_stats}a) and the overestimation of their sizes (Fig.~\ref{fig:compar_stats}b). In contrast, SAM-P’s accuracy stems from more consistent detections of both the number and size of floes. The other two methods underestimate concentration. SAM-GVF yields an error of approximately 10\%, while GVF alone performs the worst, with a discrepancy of nearly 30\%. The errors in GVF are attributed to missing floes and underestimation of their sizes, particularly in areas with inconsistent gradients across the floes.

\section{Conclusion}
\label{sec:conclusion}

Three segmentation approaches have been evaluated for analyzing close-scale optical imagery of ice floes in the winter Antarctic MIZ. The first approach employs the gradient vector flow (GVF) snake algorithm, which uses edge-based energy minimization for contour detection. The second approach utilizes the Segment Anything Model (SAM), a deep learning framework for image segmentation, applied in both its automatic mode (SAM-A) and a prompt-guided approach (SAM-P). The third approach combines SAM-P with the GVF algorithm (SAM-GVF) to leverage their complementary strengths. To ensure robust and systematic evaluation, we implemented a fully automated image processing pipeline, incorporating orthorectification and image quality enhancement to standardize images for segmentation analysis. The pipeline autonomously executes the GVF snake algorithm and strategically integrates SAM-P through automatically generated prompts directly targeting individual floes. Performance was quantitatively assessed using a benchmark dataset consisting of 20 manually segmented images representing over 2800 floes under varying lighting, weather, and ice conditions. The evaluation highlighted the relative strengths and weaknesses of each method by assessing their ability to identify and segment individual floes, estimate floe size distribution, and compute sea ice concentration. To our knowledge, this work presents the first evaluation of sea ice segmentation methods using a large-scale, manually annotated benchmark, and introduces an automated framework that enables reproducible, scalable, and robust segmentation analyses across diverse sea ice conditions.

The GVF method provided a reasonable floe size distribution due to its ability to capture floe shapes, particularly when floes were regular, circular, and medium sized, but it was less effective in detecting all floes, impacting its accuracy in estimating ice concentration. SAM-A consistently performed well on images with distinct floe edges regardless of floe shape, although its effectiveness decreased significantly in challenging environments, detecting fewer floes and providing the most divergent size distributions from the benchmark. SAM-A provided an accurate sea ice concentration estimate, albeit due to compensating factors: fewer, but larger detected floes. SAM-P was more effective at detecting floes, consistently producing accurate size distributions and robust sea ice concentration estimates across the entire sample. The SAM-GVF combination slightly improved upon SAM-P by reducing misinterpretations of complex floes, although overall performance remained similar, closely aligning with the benchmark in both floe size distribution and sea ice concentration.

SAM-P and SAM-GVF emerged as the most effective methods, demonstrating strong performance across multiple metrics, especially in accurately estimating sea ice concentration and improving floe detection. SAM-P’s computational efficiency makes it suitable for real-time applications requiring immediate results and for processing large datasets to derive statistical properties of sea ice. Conversely, SAM-GVF, despite its longer computational times, is better suited for detailed analyses where precise floe shape information is essential.

This study establishes robust methodological foundations for obtaining reliable quantitative data on sea ice floes in the MIZ from field-acquired imagery. Due to logistical and observational constraints, Antarctic sea ice datasets, especially floe size data, remain limited. By developing an effective segmentation methodology, we facilitate improved utilization of field imagery, increasing the quantity and quality of data obtained from these observational campaigns. Although SAM, as a generalist model, is not currently optimized for sea ice, our extensive dataset provides an opportunity for further training specifically for Antarctic conditions. Future research will extend this analysis to the complete image collection from multiple Antarctic expeditions and seasons, enabling comprehensive monitoring of the evolution of Antarctic sea ice.

\section*{Open Research Section}
The 20 sample images used in this study, along with the image processing scripts, are available on GitHub \cite{passerotti2025seaicefloesegmentation}.

\section*{Acknowledgments}
The expeditions were funded by the South African National Antarctic Programme through the National Research Foundation. AA and AT were partially funded by the ACE Foundation and Ferring Pharmaceuticals and the Australian Antarctic Science Program (project 4434). AT and LGB acknowledge supported by the Australia Research Council (DP200102828, DP240100325, and  LP210200927). AA acknowledges the EPSRC grant EP/Y02012X/1. MV was supported by the NRF SANAP contract UID118745. We are indebted to Captain Knowledge Bengu and the crew of the SA Agulhas II for their invaluable contribution to data collection. We acknowledge Dr L. Fascette for technical support.

\clearpage
\begin{appendices}

\section{SAM vs. SAM 2}
\label{sam2}

To evaluate how different versions of the model affect segmentation performance, we compare the outputs of SAM and the newer SAM 2 using the two representative images: one featuring distinct ice floes with well-defined edges, and another with diffuse sea ice boundaries. For each version, both automatic and prompt-based modes are tested. Manual segmentation serves as a benchmark to assess how accurately each method captures the characteristics of the floe population. A summary of the comparison is provided in Table~\ref{tableAppend}.

\begin{table}[htbp]
\centering
\caption{Comparison of SAM and SAM 2 segmentation results. For each method and image type, the table reports the number of floes (Fl.), mean diameter ($\overline{\mathrm{D}}$), ice concentration (IC), and average computational time per floe. Manual segmentation results serve as reference benchmarks.}
\label{tableAppend}
\renewcommand{\arraystretch}{1.3}
\begin{tabular}{l|cccc|cccc}
\toprule
\multirow{2}{*}[-0.25em]{Method}
 & \multicolumn{4}{c|}{\textbf{Distinct edges image}}
 & \multicolumn{4}{c}{\textbf{Diffuse boundaries image}}\\
\cmidrule(lr){2-9}
 & \raisebox{1em}{Fl.} & \raisebox{1em}{$\overline{\mathrm{D}}$} & \raisebox{1em}{IC} & \shortstack{Average\\CPU time\\per floe}
 & \raisebox{1em}{Fl.} & \raisebox{1em}{$\overline{\mathrm{D}}$} & \raisebox{1em}{IC} & \shortstack{Average\\CPU time\\per floe}\\
\midrule
Manual
& 196 & 3.88\,m & 57\% & N/A
& 250 & 3.05\,m & 55\% & N/A\\
SAM-A
& 168 & 4.20\,m & 55\% & 0.8\,s
& 166 & 3.9\,m & 51\% & 0.8\,s\\
SAM-P
& 192 & 3.88\,m & 55\% & 0.3\,s
& 208 & 3.44\,m & 53\% & 0.3\,s\\
SAM2-A
& 112 & 4.47\,m & 40\% & 0.7\,s
& 86 & 4.79\,m & 37\% & 0.9\,s\\
SAM2-P
& 175 & 3.93\,m & 50\% & 0.3\,s
& 172 & 3.74\,m & 49\% & 0.3\,s\\
\bottomrule
\end{tabular}
\end{table}

SAM 2 consistently under performs relative to the original SAM across all evaluated metrics. In automatic mode, it detects 33\% fewer floes than SAM in the image with distinct edges, and 48\% fewer in the image with diffuse boundaries. In prompt-based mode, the reductions are 9\% and 17\%, respectively.

SAM 2 tends to overestimate floe size, with mean diameters 15\% (distinct edges) and 57\% (diffuse boundaries) larger than the manual benchmark in automatic mode, and 1\% and 23\% larger, respectively, in prompt-based mode. In comparison, SAM in automatic mode overestimates floe size by only 8\% (distinct edges) and 28\% (diffuse boundaries), while its prompt-based mode exactly matches the benchmark in the distinct edges image and shows a modest 12\% increase in the diffuse boundaries image.

Ice concentration from SAM 2 is generally lower than that from SAM, especially in the diffuse boundaries image. The average ice concentration errors relative to the manual benchmark are 18\% in automatic mode and 7\% in prompt-based mode for SAM 2, compared to just 3\% and 2\%, respectively, for SAM. Computational time per floe remains similar between SAM and SAM 2.

\section{Benchmark dataset photos}
\label{benchdata}

\begin{figure}
\centering\includegraphics[width=0.85\textwidth]{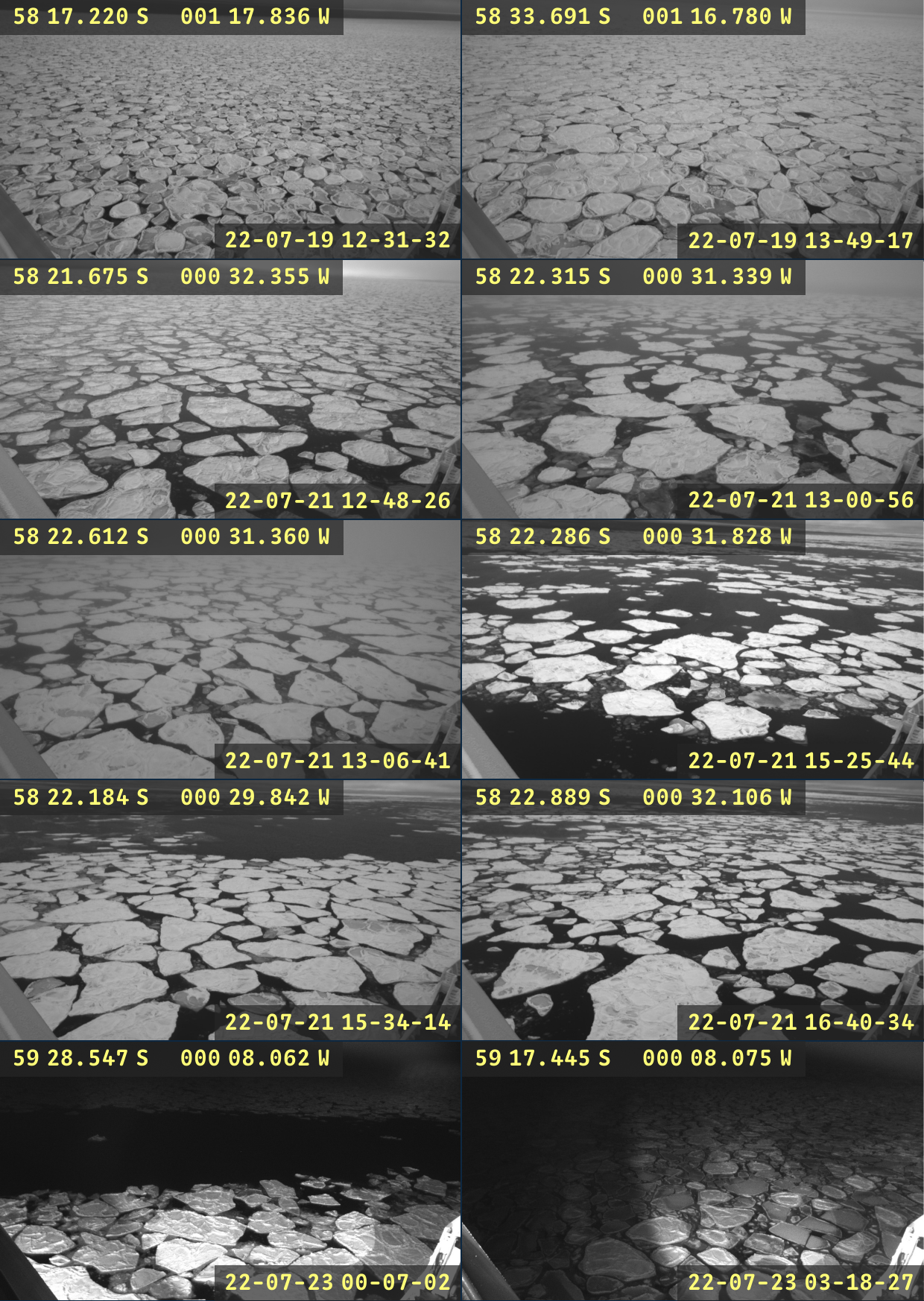}
\caption{Ten original photos from the benchmark dataset (out of twenty total), each annotated with timestamps and approximate GPS coordinates.}
\label{fig:dataset1}
\end{figure}

\begin{figure}
\centering\includegraphics[width=0.85\textwidth]{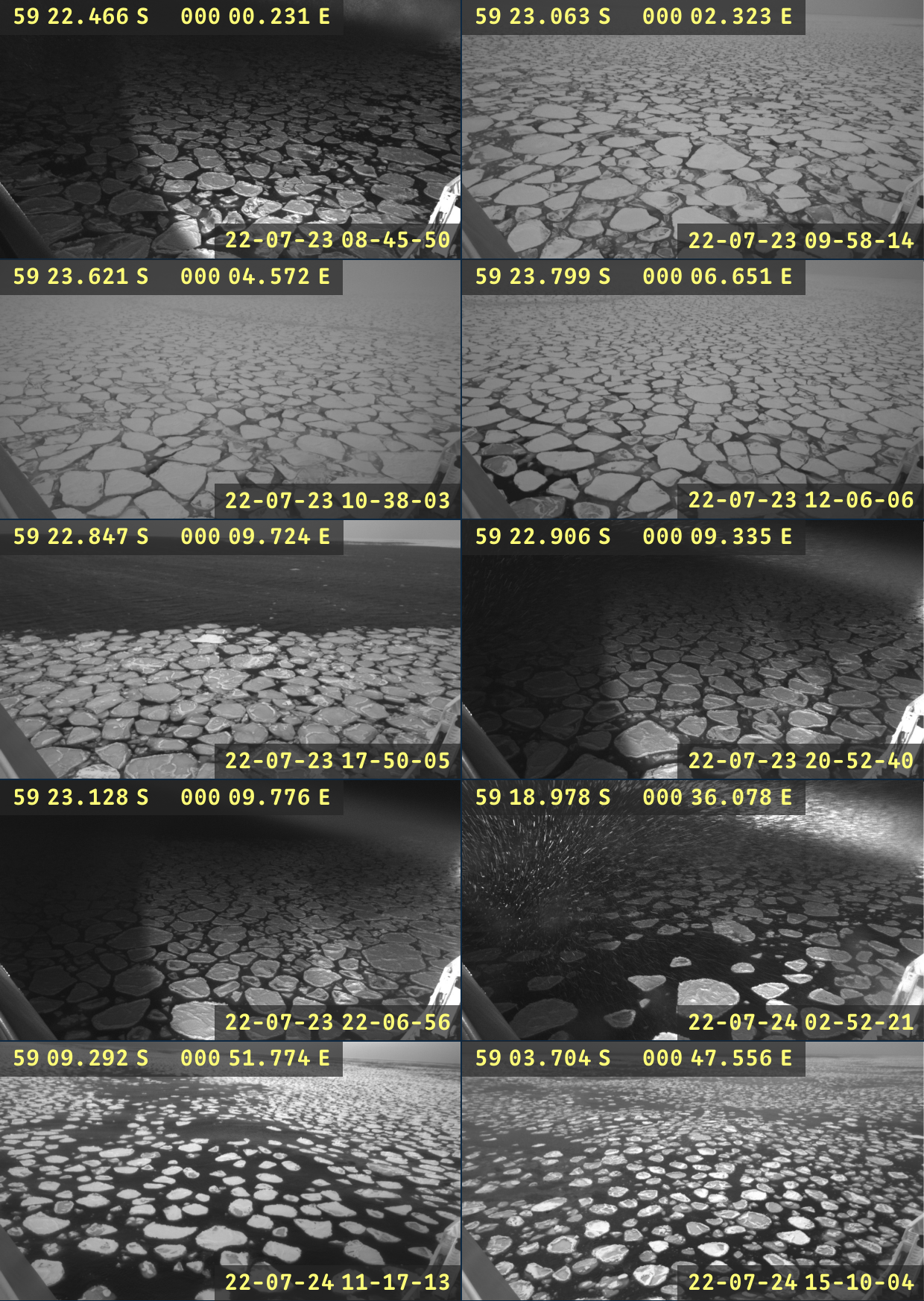}
\caption{The remaining ten original photos from the benchmark dataset, each annotated with timestamps and approximate GPS coordinates.}
\label{fig:dataset2}
\end{figure}

Figures~\ref{fig:dataset1} and \ref{fig:dataset2} present the twenty photos comprising the benchmark dataset, illustrating a range of ice conditions and lighting environments. Approximately 30\% of the dataset represents optimal conditions, featuring distinct ice floes with clearly defined edges, while the remaining 70\% captures more complex scenarios characterized by diffuse sea ice boundaries. Additionally, roughly half of the benchmark dataset consists of floes with round or circular shapes, while the other half features floes that are more angular, squared, and less regular in shape.

\end{appendices}
\clearpage

\bibliographystyle{unsrtnat}

\end{document}